\documentclass[letterpaper,twocolumn,10pt]{article}

\usepackage{lipsum}

\newcommand\blfootnote[1]{%
  \begingroup
  \renewcommand\thefootnote{}\footnote{#1}%
  \addtocounter{footnote}{-1}%
  \endgroup
}

\usepackage[ruled,vlined]{algorithm2e}
\usepackage{algpseudocode}% http://ctan.org/pkg/algorithmicx

\usepackage{mathtools}

\usepackage{mathtools, nccmath}
\usepackage[b]{esvect} 

    \newcommand\MTkillspecial[1]{% helper macro
    \bgroup
    \catcode`\&=9
    \let\\\relax%
    \scantokens{#1}%
    \egroup
    }
    \DeclarePairedDelimiter\parens()
    \reDeclarePairedDelimiterInnerWrapper\parens{star}{
    \mathopen{#1\vphantom{\MTkillspecial{#2}}\kern-\nulldelimiterspace\right.}
    #2
    \mathclose{\left.\kern-\nulldelimiterspace\vphantom{\MTkillspecial{#2}}#3}}

\usepackage{graphicx}
\graphicspath{{figures/}}
\DeclareGraphicsExtensions{.jpg,.png}
\usepackage{etoolbox}

\usepackage[caption=false,font=footnotesize]{subfig}

\usepackage{booktabs}

\usepackage{multirow,makecell}%"makecell" is for Xcline

\usepackage{framed}
\usepackage{float}

\usepackage{listings}
\usepackage{comment}
\usepackage{array}
\usepackage{rotating}
\usepackage{adjustbox}
\usepackage{tabularx}
\usepackage[framemethod=default]{mdframed}
\usepackage{tikz}
\usepackage{pgfplots, pgfplotstable}
\usetikzlibrary{patterns}

\usepackage[T1]{fontenc}

\usepackage{pgfplotstable}
\usepackage{filecontents}

\usepackage{array}
\usepackage{xcolor}
\usepackage{colortbl}

\PassOptionsToPackage{hyphens}{url}
\usepackage[hidelinks]{hyperref}
\usepackage{adjustbox}
\usepackage{fancyvrb}
\usepackage{balance}
\usepackage{xspace}
\newcommand{\etal}{\textit{et al.}\xspace}
\newcommand{\etc}{\textit{etc.}\xspace}
\newcommand{\ie}{\textit{i.e.,}\xspace}
\newcommand{\eg}{\textit{e.g.,}\xspace}
\newcommand{\cf}{\textit{cf.}\xspace}

\newcounter{protocol}[section]

\newcolumntype{B}[2]{%
    >{\adjustbox{angle=#1,lap=\width-(#2)}\bgroup}%
    l%
    <{\egroup}%
}
\newcolumntype{Z}[2]{%
    >{\adjustbox{angle=#1,lap=\width-(#2)}\bgroup}%
    l%
    <{\egroup}%
}

\newcolumntype{L}[1]{>{\raggedright\let\newline\\\arraybackslash\hspace{0pt}}m{#1}}
\newcolumntype{C}[1]{>{\centering\let\newline\\\arraybackslash\hspace{0pt}}m{#1}}
\newcolumntype{R}[1]{>{\raggedleft\let\newline\\\arraybackslash\hspace{0pt}}m{#1}}

\definecolor{semi-light-gray}{gray}{0.7}
\definecolor{light-gray}{gray}{0.8}

\usepackage{pgfplotstable}
\usepgfplotslibrary{statistics}

\newenvironment{customlegend}[1][]{%
    \begingroup
    \csname pgfplots@init@cleared@structures\endcsname
    \pgfplotsset{#1}%
}{%
    \csname pgfplots@createlegend\endcsname
    \endgroup
}%
\def\addlegendimage{\csname pgfplots@addlegendimage\endcsname}

\newcommand{\wdthheat}{1.8in}
\newcommand{\hghtheat}{1.8in}

\newcommand{\hghtthird}{1.89in}
\newcommand{\wdthval}{1.5in}
\newcommand{\hghtval}{1.5in}
\newcommand{\wdth}{2.0in}
\newcommand{\hght}{2.0in}
\newcommand{\spcbtwn}{-18 pt}

\usepackage{usenix2019_v3,epsfig,endnotes}
\usepackage{xcolor}
\usepackage{soul}
\begin{document}

%don't want date printed
%\date{}

%make title bold and 14 pt font (Latex default is non-bold, 16 pt)
%\title{Detecting Distance Enlargement Attacks in Ultra-Wideband}
\title{UWB-ED: Distance Enlargement Attack Detection in Ultra-Wideband}

\author{\rm Mridula Singh, Patrick Leu, AbdelRahman Abdou, Srdjan Capkun \\
	Dept. of Computer Science\\
	ETH Zurich\\
  \{firstname.lastname\}@inf.ethz.ch
}

\maketitle

% Use the following at camera-ready time to suppress page numbers.
% Comment it out when you first submit the paper for review.
%\thispagestyle{empty}
\pagenumbering{gobble}

\subsection*{Abstract}
Mobile autonomous systems, robots, and cyber-physical systems rely on accurate positioning information.  To conduct distance-measurement, two devices exchange signals and, knowing these signals propagate at the speed of light, the time of arrival is used for distance estimations. Existing distance-measurement techniques are incapable of protecting against adversarial distance enlargement---a highly devastating tactic in which the adversary reissues a delayed version of the signals transmitted between devices, after distorting the authentic signal to prevent the receiver from identifying it. The adversary need not break crypto, nor compromise any upper-layer security protocols for mounting this attack. No known solution currently exists to protect against distance enlargement. We present \textit{Ultra-Wideband Enlargement Detection} (UWB-ED), a new modulation technique to detect distance enlargement attacks, and securely verify distances between two mutually trusted devices. We analyze UWB-ED under an adversary that injects signals to block/modify authentic signals. We show how UWB-ED is a good candidate for 802.15.4z Low Rate Pulse and the 5G standard.

\newcommand{\first}{Attack Plausibility}
\newcommand{\second}{Robust Code Verification}

\blfootnote{Version: \today.}

\section{Introduction}

\begin{figure}
	\centering
	\includegraphics[width=1 \linewidth]{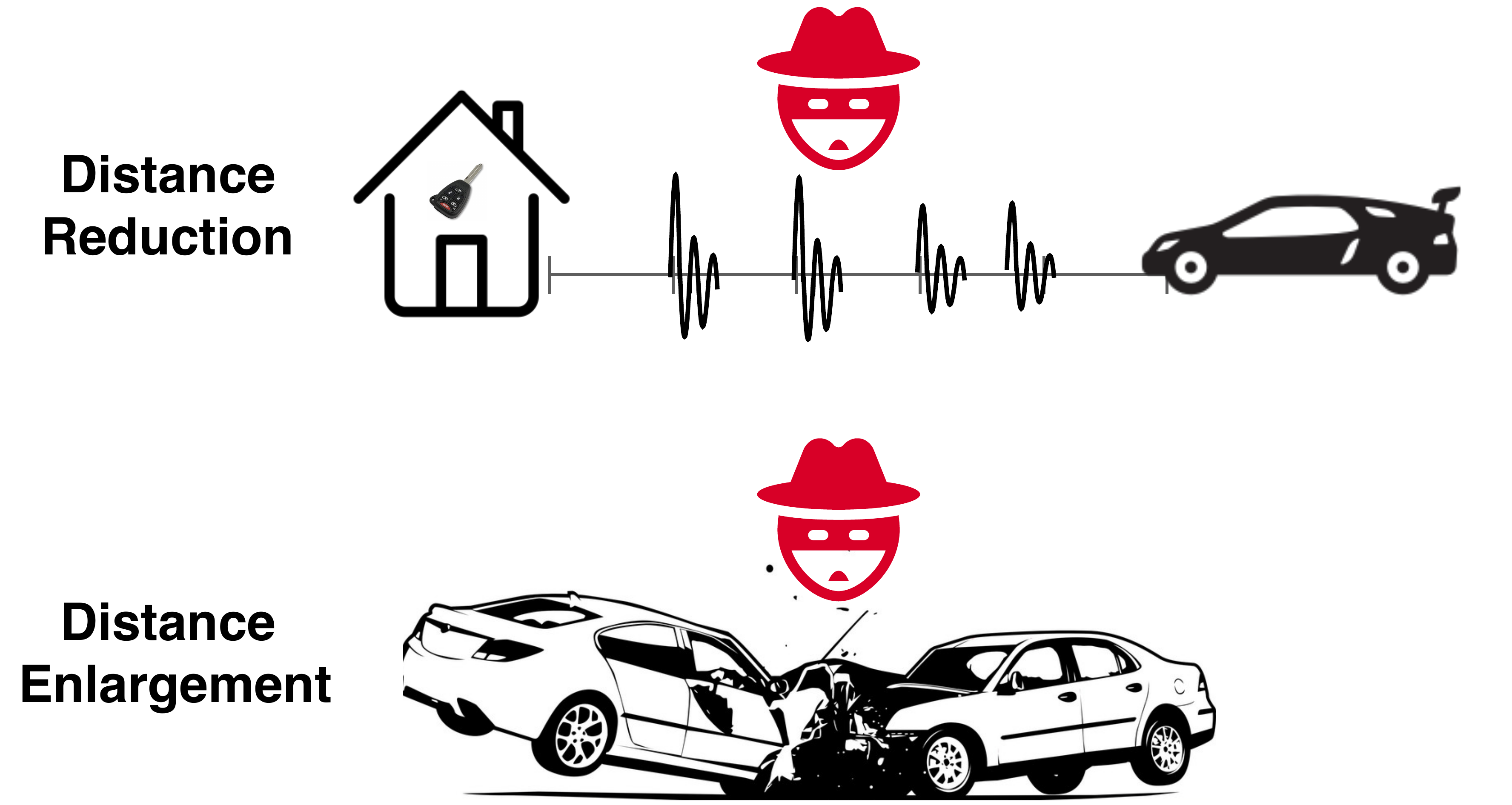}
	\caption{Ranging systems are vulnerable to distance reduction and enlargement attacks.}
	\label{fig:attacks}
\end{figure}

%Why ranging is important
Ranging and positioning information is often necessary for mobile autonomous systems, robots and cyber-physical systems to operate successfully. These systems are used in security and safety critical applications. Drones are becoming more popular for transportation and rescue~\cite{postdrones}, and autonomous systems are being increasingly tested and integrated as part of the ecosystem. The 5G community emphasizes the importance of designing the wireless protocols for the safety of the autonomous vehicles~\cite{5G_Vehicular_Networks_Positioning}. A stringent requirement for these systems is to avoid crashing into, \eg buildings, pedestrians, properties, or each other~\cite{bbcdrones}. For example, keeping drones and autonomous vehicles on their intended paths and preventing their collision can be achieved only if they are able to calculate their relative positions accurately and securely. Figure \ref{fig:attacks} shows that an adversary can manipulate the perceived distance between two mutually trusted devices by the distance reduction and enlargement attacks. 

Conventional ranging systems, such as GPS and WiFi Positioning Systems (WPS)~\cite{zandbergen2009accuracy}, are useful for benign environments and coarse-granular geolocation. However, they provide insufficient precision for accurate distance estimations (\eg $cm$-level granularity), suffer availability constraints (\eg indoors, outdoors), and are relatively slow to calculate locations for fast and mobile autonomous systems. More importantly, the aforementioned ranging systems are susceptible to various spoofing attacks~\cite{GPS_Spoofing1,Spoofing2,Nils_WLAN_attack}.

Two-way time-of-flight (ToF)-based ranging systems (which map ToF to distance as signals propagate at the speed of light) have the potential to conduct accurate, fast, and secure distance measurements. Examples include high precision Ultra-wide Band~(UWB) ranging systems, some of which are now available off-the-shelf~\cite{timedomain,3db,deca,zebra}. Numerous previous efforts were directed towards protecting these systems from distance-reduction attacks, \eg for access control. These mainly rely on the principle that propagation speeds are bounded by the physical characteristics of the media, and cannot be sped-up. For example, distance bounding protocols return an upper bound on the measured distance, armed by the fact that an adversary would not succeed in guessing (secret) bit level information~\cite{Brands1994,DB_EPFL}. Other techniques are based on tailoring modulations to prevent distance-reduction attacks at the physical layer~\cite{mridula_eprint_UWB_PR}. None of these approaches prevent distance enlargement attacks. 

Distance enlargement attacks can deviate vehicles from their intended paths, or cause physical collisions. Existing protection approaches rely on dense, and often fixed, verification infrastructures, \eg towers. These may not exist, and often do not; installing them in outdoor settings is a costly affair, and not necessarily feasible (\eg in drone-based military missions behind enemy lines). 
Distance enlargement is a more devastating attack than distance shortening because an adversary in the communication range only needs to annihilate (cancel)~\cite{Cristina_ESORICS} or distort the authentic signals to prevent the receiver from identifying them and using their time-of-arrival (ToA) for ranging. The adversary then simply replays a delayed version of the authentic signals, which it has already received by positioning itself in the vicinity of the sender or the receiver. The adversary need not guess these signals, nor compromise any upper-layer protocols to do that. The amount of delay corresponds to the adversary-intended distance to enlarge. In a collision-avoidance system of automobiles or self-driving cars for example, a few meters ($\sim$ a few nanoseconds) could be catastrophic.

We present \textit{Ultra-Wideband Enlargement Detection} (UWB-ED)---the first known modulation technique to detect distance enlargement attacks against UWB ranging based on ToF. UWB-ED relies on the interleaving of pulses of different phases and empty pulse slots (\ie on-off keying). 
Unable to perfectly guess the phase, this leaves the adversary with a 50\% chance of annihilating pulses (similarly for amplification). As a result, some of the affected (authentic) pulses will be amplified, while others will be annihilated. Unaffected pulses will remain intact, while positions that originally had no pulses may now have adversary-injected ones. The technique presented herein gets the receiver to seek evidence indicating whether such a deformed trail of pulses in the transmission was indeed authentic, albeit corrupt.

Similar to Singh \etal~\cite{mridula_eprint_UWB_PR} (which addresses distance-reduction attacks), we leverage a randomized permutation of pulses. However, unlike~\cite{mridula_eprint_UWB_PR}, we cannot simply look for whether these are out of order, and ignore them if so because that is precisely the adversary's objective in distance-enlargement: misleading the receiver to ignore the authentic signals. Instead, UWB-ED checks the \emph{energy distribution} of pulses: comparing the aggregate energies of a subset of pulses at the positions where high energy was expected (as per the sender-receiver secret pulse-permutation agreement), with others where low energy was expected. To subvert this, the adversary would be forced to inject excessive energy throughout the whole transmission, which could then be detected using standard DoS/jamming-detection techniques.

We derive the probability that an adversary succeeds in a distance-enlargement attack against UWB-ED. This is also useful in setting input parameters, \eg balancing an application's security requirements and ranging rate, while accounting for channel conditions. For example, we show how proper parameterization of UWB-ED limits an adversary's success probability in enlarging distances to $< 0.16\times 10^{-3}$. 

In summary, the paper's contributions are twofold. 
\begin{itemize}
	\item  UWB-ED---a novel, readily-deployable modulation technique for detecting distance enlargement attacks against UWB ToF ranging systems, requiring absolutely no verification infrastructure, and making no impractical assumptions limiting adversarial capabilities. 
	\item  Analytical evaluation to UWB-ED, where the probability of adversarial success is derived as a function of input parameters and channel conditions. This evaluation is also validated using simulations.
\end{itemize}

The sequel is organized as follows. Sections~\ref{sec:background} and \ref{sec:threatmodel} provide background and detail the threat model. The new distance enlargement detection technique is explained in Section~\ref{sec:proposedsystem}, and evaluated in \ref{sec:eval}. Section~\ref{sec:security} complements with a related discussion, and \ref{sec:relatedwork} is related work. Section~\ref{sec:conc} concludes.

\section{Background and Motivation}
\label{sec:background}

%infrastructure
A device's position can be estimated using the distances between itself and other landmarks with known locations; or it could be expressed using a coordinate system, \eg in a Cartesian plane. The distance between two devices can be measured using radio signal properties, such as received signal strength~\cite{RADAR_RSSI}, phase \cite{DinaKatabiPHASE}, or the signal's propagation time including ToF and ToA~\cite{SurePoint}. Reduction or enlargement of the calculated distances can lead to wrong positioning. 

Adversarial distance reduction has been analyzed in previous literature~\cite{VerifiableMultilateration}, but limited work was performed on enlargement attacks. Preventing enlargement is achieved when a node is inside a polygon determined by an infrastructure of devices/towers, where verifiable multilateration~\cite{VerifiableMultilateration} is applied. Enlargement attacks are harder to detect without an infrastructure. Signal strength-based systems do not provide strong security guarantees during high variations of signal strengths in some channel conditions. For distance reduction attacks, the adversary can amplify a degraded signal but for enlargement, degradation is in the adversary's favor.

One-way ToF systems, such as GPS, can be spoofed to reduce/enlarge distances~\cite{GPS_Spoofing1,Spoofing2}. Two-way ToF, such as UWB, provides secure upper bound by using distance bounding along with secure modulation techniques~\cite{Brands1994,DB_EPFL,mridula_eprint_UWB_PR}. This provides strong guarantees against reduction attacks, but is still susceptible to enlargement attacks.

\subsection{UWB}   

IEEE 802.15.4a and IEEE 802.15.4f have standardized impulse radio UWB as the most prominent technique for precision ranging. IEEE 802.15.4z~\cite{802.15.4z} is in the process of standardizing UWB to prevent attacks on the ranging systems. Off-the-shelf UWB ranging systems were recently developed~\cite{timedomain,3db,deca,zebra}, and the research community/industry has expressed tremendous interest in these systems (\eg for autonomous vehicles). Because current standards do not prevent enlargement attacks, it is important to mitigate them before standards are deployed in practice.

\paragraph{Symbol Structure.}
\label{sec:symbol_structure}

UWB systems operate over wide segments of licensed spectrum. They have to be compliant with stringent regulatory constraints. Firstly, the power spectral density should not exceed $-41.3$dBm/MHz, averaged over a time interval of 1ms. Secondly, the power measured in a 50MHz-bandwidth around the peak frequency is limited to 0dBm. Due to these constraints, the power per pulse is limited. To support longer distances, the energy of multiple pulses is aggregated to construct meaningful information. 
Figure~\ref{fig:enlargementattack} shows On-Off-keying (OOK) modulation, as used in IEEE 802.15.4f-based UWB ranging systems. Each symbol has two pulses and two empty slots. The symbol length is represented as $T_b$ and the spacing between consecutive pulses is $T_s$. Information bits are encoded in the position of the pulse.%The enlargement attack on such system is discussed further in Section~\ref{sec:attackdesc}.}

\paragraph{Symbol Detection.}   % in non-Coherent receiver}
\label{sec:symbol_receiver}

\begin{figure}
	\centering
	\includegraphics[width=1\linewidth]{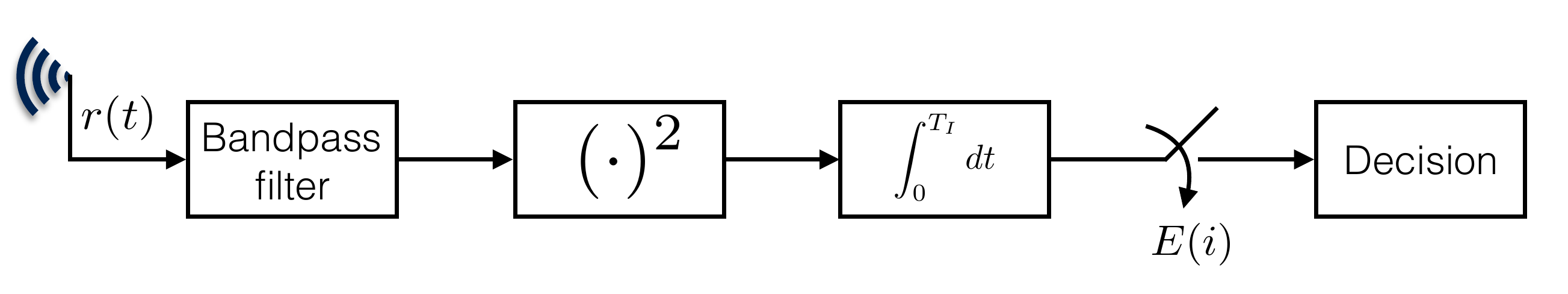}
	\caption{Non-coherent energy detector receiver.}
	\label{fig:noncoherentreceiver}
\end{figure}

Figure~\ref{fig:noncoherentreceiver} shows a conventional non-coherent energy detector~(ED) receiver \cite{Non-Coherent-receiver-explain}. The energy detector receiver is consist of square-law device to compute instantaneous received signal power and an energy integrator. For the received signal $r(t)$, the output of the receiver can be expressed as:
\begin{equation}
E(k) = \int_{ T_s*k }^{T_s*k + T_I} [r(t)]^2 dt
\end{equation}
where $T_s*k$ is the integration start time, $T_I$ the integration window size, and $T_s$ the spacing between consecutive pulses. 

These receivers perform squaring and integration, making phase information irrelevant for pulse detection. In the case of multi-pulse per symbol, the energies of multiple pulses are aggregated. For the orthogonal hypothesis tests $H_1$ and $H_0$ for bit 1 and 0 respectively, the decision of the ED receiver is made in favor of the positions with higher energy.

\begin{equation}
b(i) =
\begin{cases}
0  &  E_{H_0}(i) \ge  E_{H_1}(i)\\
1      & E_{H_0}(i) <  E_{H_1}(i) 
\end{cases}
\end{equation}

\subsection{Distance-Enlargement Attack}
\label{sec:attackdesc}

\begin{figure}
	\centering
	\includegraphics[width=1\linewidth]{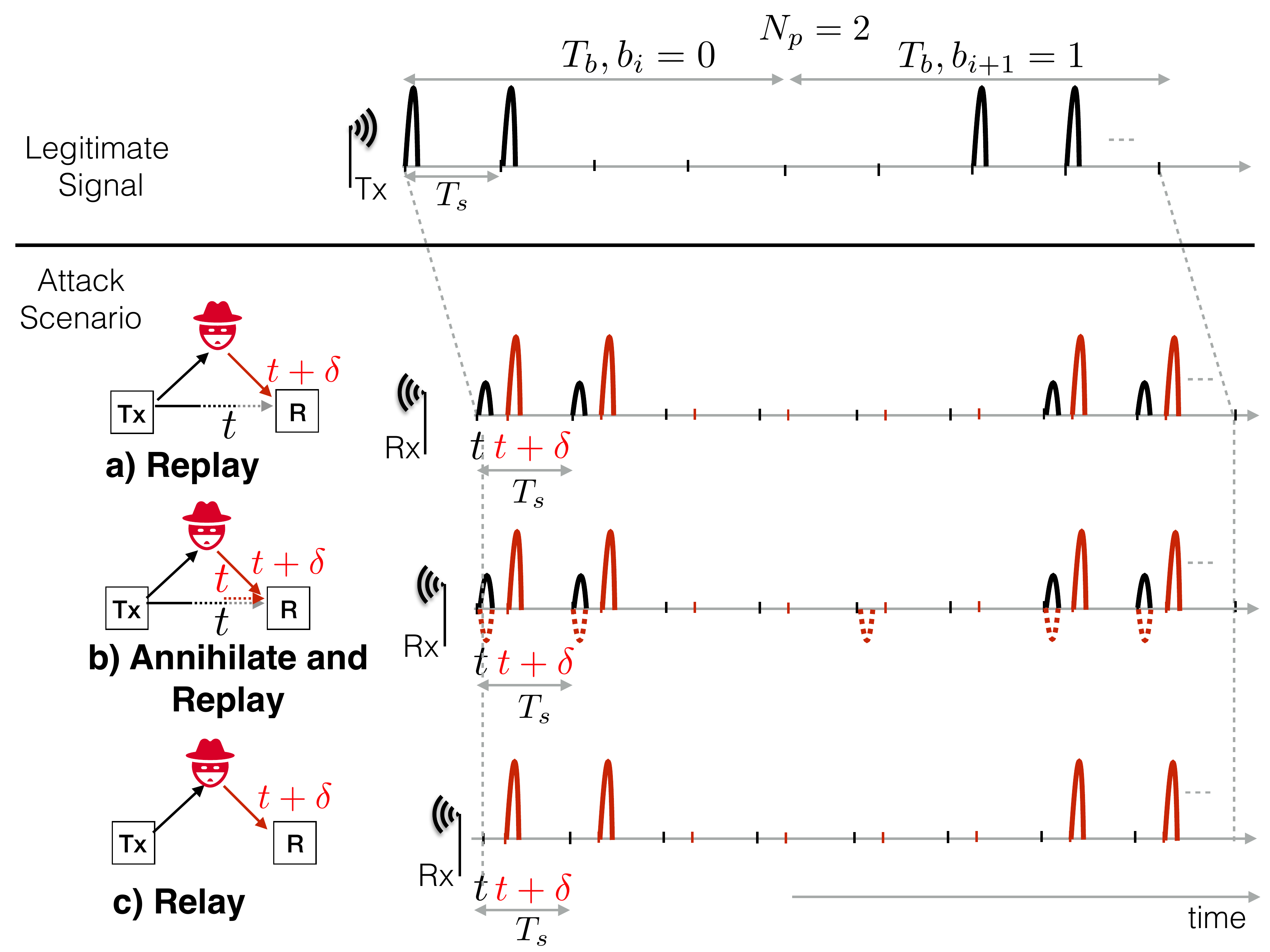}
	%The adversary cannot control the overshadowing effect because the receiver still recovers authentic signals. 
	\caption{Various attack scenarios on UWB.Black and red colors represent authentic and adversary signals respectively. Dotted red represent adversarial signal-annihilation attempts.}
	\label{fig:enlargementattack}
\end{figure}

In contrast to reduction attacks, to enlarge the distance, the adversary need not predict the authentic signal. Instead, it replays the authentic signal by replaying an amplified version of it after some delay. The receiver gets both, authentic and adversary's signal superimposed. Because these authentic signals also reach the receiver, the adversary cannot control how the receiver processes them. 
None of the existing ranging systems is secure against enlargement attack- be it UWB -802.15.4z, WiFi- 802.11, or GPS. Signal replay is  a typical strategy to mount distance enlargement attacks. Other enlargement attacks, such as jamming, alters the output of the receiver's automatic gain control~(AGC), and are likely to expose the adversary
%by enforcing the limit on received power
~\cite{enlargement_miscontrol,Taponecco_Overshadow}.
%power to overshadow legitimate signal, by jamming or annihilating legitimate signal and then injecting his signal \cite{Cristina_ESORICS,Nils_WLAN_attack}. 
Complementing signal replay by signal annihilation prevents the receiver from detecting the authentic signal. Annihilation is possible due to the predictable symbol structure. 

\begin{figure}
	\centering
	\includegraphics[width=1\linewidth]{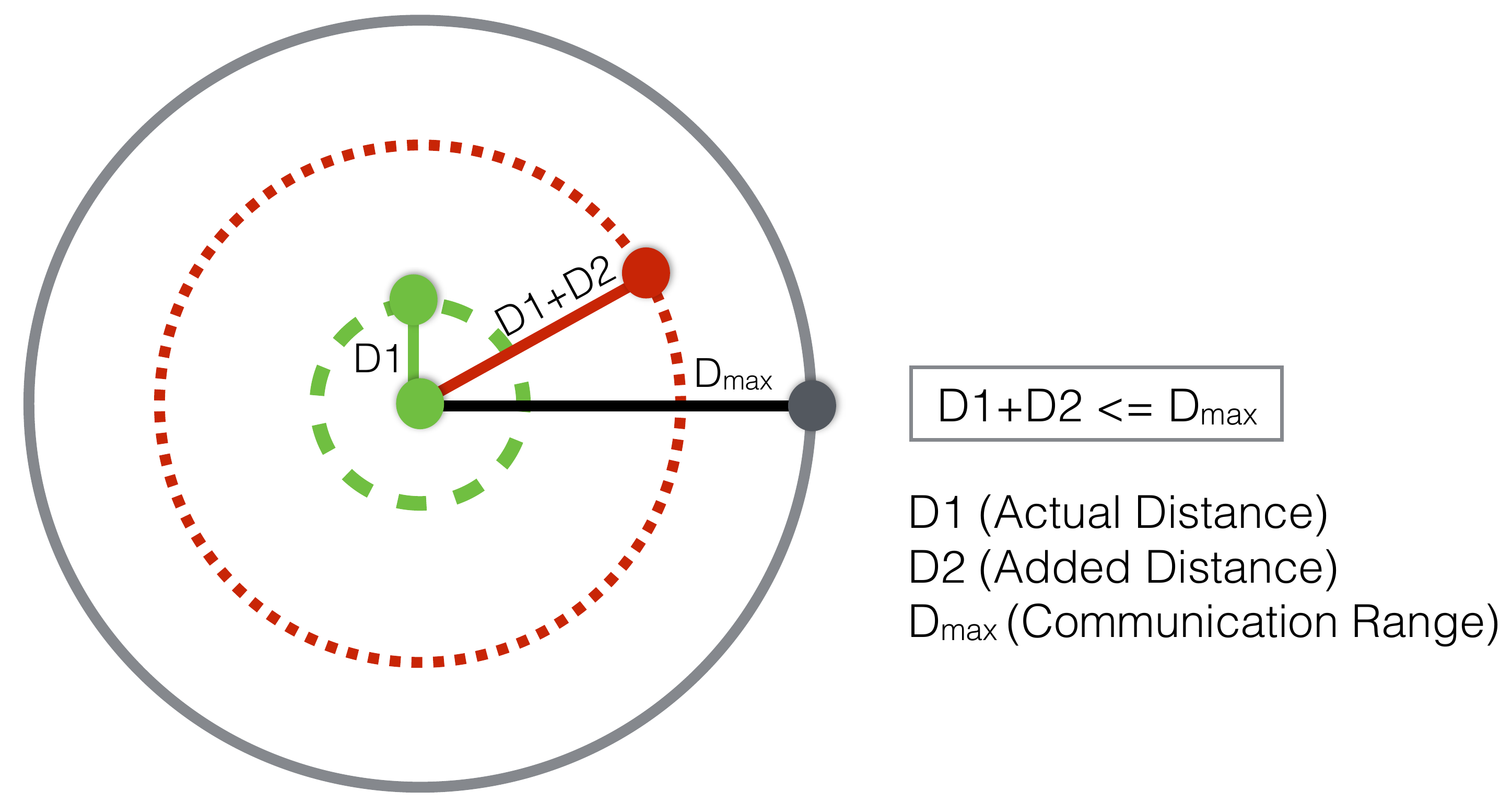}
	\caption{If $D1+D2>D_\text{max}$, the devices realize they are outside each other's communication range without the need to run distance-enlargement detection protocol.}
	\label{fig:d1d2range}
\end{figure}

In Fig.~\ref{fig:enlargementattack}, the devices know each other's communication range, and could verify that they are within that range, \eg using secure ranging (see Fig.~\ref{fig:d1d2range}). For short LoS distances, a symbol length of $N_p=1$ (\ie one pulse-per-symbol) could suffice. Longer distances are attained by longer symbols ($N_p=2$ in Fig.~\ref{fig:enlargementattack}). Pulses are separated by time $T_s$, which should be more than the channel's delay spread. The length of the symbol $(T_b)$ is determined by the number of pulses per symbol, and the interval between two consecutive pulses $(N_p \cdot T_s)$. Figure~\ref{fig:enlargementattack} also shows instances of replay attacks on these symbols. When an adversary replays authentic signals after some delay ($\delta$), both authentic and replayed signals are received. To deceive the receiver, the adversary needs to annihilate authentic signals. 

In Fig.~\ref{fig:enlargementattack}a, an authentic signal reaches the receiver at time $t$, and the adversary's signal at $t+\delta$. %, where $t+\delta$ corresponds to the adversary-influenced distance at the commitment phase. 
If the receiver backtracks in time (searching for earlier-received signals), the authentic signal will be encountered. Figure~\ref{fig:enlargementattack}b shows how the predictability of the symbol structure enables an adversary to annihilate its pulses (by emitting a reciprocal pulse phase), preventing the receives from detecting it. Figure~\ref{fig:enlargementattack}c shows the case when nodes are not in the communication range (or signal is attenuated by channel condition); the receiver does not get authentic signals, just adversary-relayed (and delayed) signals.

\section{Threat Model}
\label{sec:threatmodel}

We focus on the scenario where there are two devices in a wireless network that are interested to securely measure the physical distance between them, and protect the measurements from a third-party adversary. The devices know their maximum communication range. The adversary's objective is to enlarge the distance that the devices measure. The adversary cannot directly block or modify messages on the channel (\cf~Dolev-Yao's adversary \cite{DolevYao}); it can rather inject signals, and through such injection it can block/modify the authentic signals. If successful, this injection can lead to jamming, signal annihilation, and/or content modification.  
This model captures the capabilities of man-in-the-middle (MITM) attacks in wireless settings, and is typical in previous literature~\cite{ICodes,ICodes_Gollakota}. The model also fits well with our target application scenario: the communicating devices are typically mobile and move (drive or fly) in formation. In such scenarios, it is unlikely that an adversary prevents the signals of one device from reaching the other by physical obstacles, and is thus limited to injecting signals.

We assume the adversary is able to communicate and listen on any channel the devices use. However, because the devices are communicating over UWB, the adversary is unable to deterministically annihilate pulses without knowing their phase (positive or negative). Existing hardware is not fast enough to enable the adversary to sample a pulse's phase and react by injecting the reciprocal pulse promptly due to the very narrow UWB pulse width of $\approx 2\ ns$. We therefore assume that the adversary will not be able to deterministically annihilate pulses from the channel, only with some probability $<1$. It succeeds in annihilating pulses if it guesses the phase of the pulse correctly. We over-approximate the adversary by providing the capability to synchronize attack signal with the authentic transmission. Signal synchronization is a hard problem, but an adversary can achieve it by using stable clock and distance information. 
We assume the adversary knows the actual physical distance between the two devices at any point in time. The adversary can calculate this using several means, \eg by eavesdropping on unencrypted position announcements the devices make. The adversary can also position itself along the direct path between the two devices, measure the distance between itself and each from that position, and add both distances. To measure these distances, the adversary's device can perform two-way ranging with each device independently, pretending to be the other device; or even without such impersonation, it could perform one-way ranging after synchronizing its clock with each device separately.

We assume the devices themselves are not compromised; the adversary cannot attach a physical cable to their interfaces, nor hijack their firmware. However, the adversary can have multiple network cards and antennas, and is not energy-bounded. It can be stationary or mobile. 

UWB-ED (Section~\ref{sec:proposedsystem}) involves transmitting, between the victim devices, a code of $n$ pulses, $\alpha$ of which are data-representing, and the remaining $\beta$ are absent of energy, where $n=\alpha+\beta$. We assume the adversary knows the values of $\alpha$ and $\beta$, but not the positions of these pulses in the transmission. (Their positions are determined by both devices pseudo-randomly in each transmission.) The adversary can learn these parameters by remaining passive in the vicinity of the victim devices, silently observing their transmissions.

Finally, we assume that it is not in the adversary's interest to prevent the devices from communicating, \eg by shielding them, or jamming the channel.

%!TEX root =  ../main.tex

\section{UWB-ED Design}
\label{sec:proposedsystem}

\begin{figure}
\centering
\includegraphics[width=1 \linewidth]{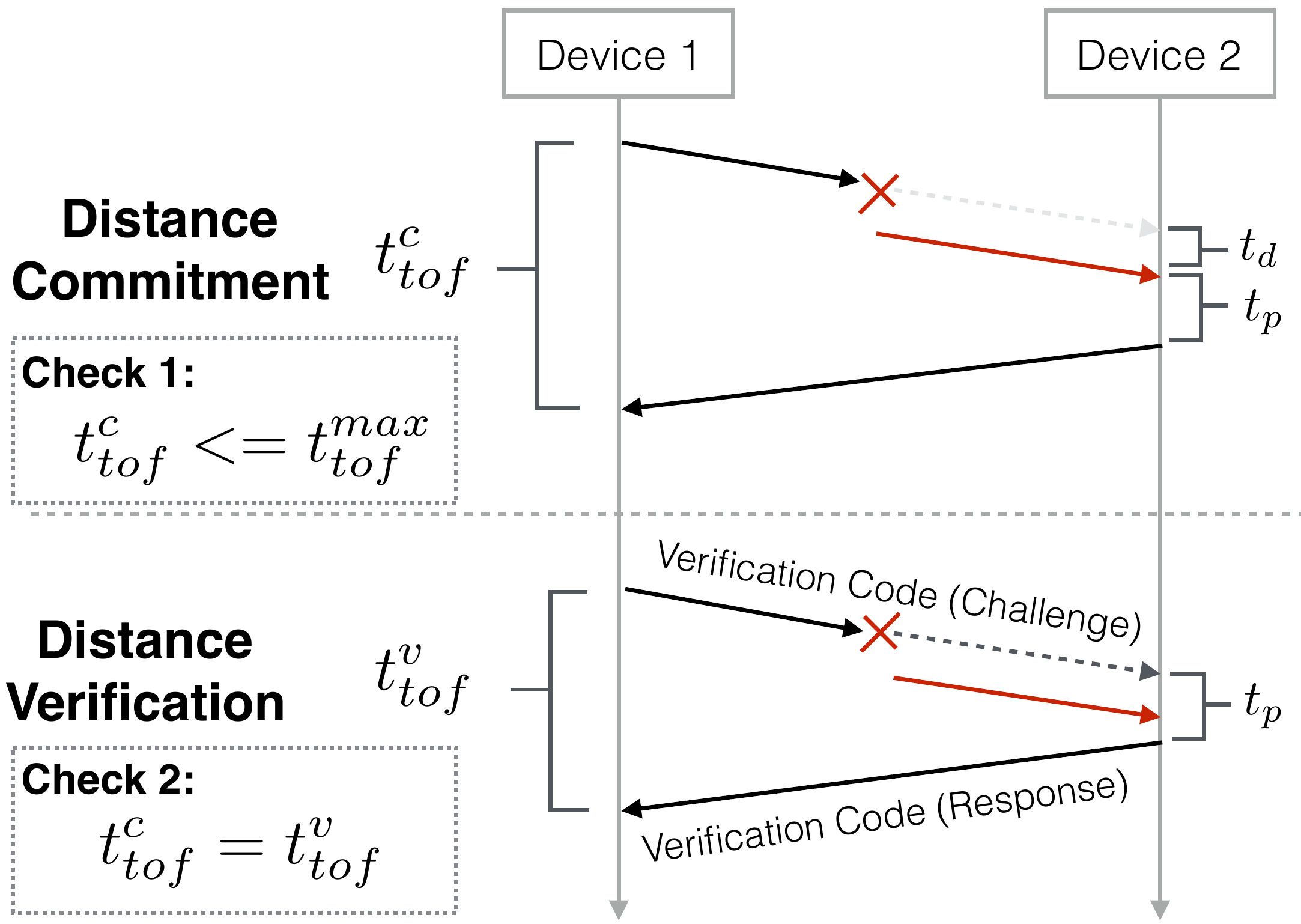}
\caption{Timing diagram of UWB-ED operation. See inline (Section~\ref{sec:proposedsystem}) for notation.}
\label{fig:protocolflow}
\end{figure}

UWB-ED consists of two phases conducted between both devices: Distance Commitment and Distance Verification. Figure~\ref{fig:protocolflow} shows a timing diagram of both phases. In the first, the devices measure the distance between them using a two-way ranging protocol. The distance measured in this phase ($t^c_{tof}$) should not exceed the supported communication range ($t^{max}_{tof}$). In the distance verification phase, the devices measure their distance by exchanging verification codes (generated using a special UWB-ED modulation). To detect enlargement attacks, devices look for distorted traces of that code. The attack is detected when such traces are found, $t^c_{tof} > t^{max}_{tof}$, or when $t^c_{tof}\neq t^v_{tof}$ (Fig.~\ref{fig:protocolflow}). By enlarging distance in the commitment phase, the adversary increases $t^c_{tof}$ by $t_d$, but fails to enlarge the distance in the verification phase. Annihilation attempts on the challenge frame are shown, but the adversary can also attack responses from both devices.

{\bf Distance Commitment Phase.} The devices measure secure upper bound by using distance bounding along with secure modulation techniques~\cite{Brands1994,DB_EPFL,mridula_eprint_UWB_PR}. This provides strong guarantees against reduction attacks but is susceptible to enlargement attacks. The distance committed in this phase should not exceed the communication range (\ie an enlargement attack is detected when $t^c_{tof} > t^{max}_{tof}$). This check ensures that the nodes can communicate without a relay. An adversary enlarging distance by more than the communication range is also exposed using this check. 

{\bf Distance Verification Phase.} In this phase, the committed distance is verified, \ie an enlargement attack is detected when $t^c_{tof} \ne t^v_{tof}$. To achieve this, the devices measure their distance using round-trip time-of-flight, with both challenge and response messages protected using specially crafted \emph{verification codes} (\ie special UWB-ED modulation). In this exchange, the sender initiates the distance verification phase by transmitting a verification code; the receiver tries to detect the presence of that code, or traces thereof, in the transmission, despite the adversary's efforts to \emph{trail-hide} its existence from the channel (Section~\ref{sec:attackdesc}). The verification code and its check is applied to both time-of-flight messages. Both devices first agree on the code's structure as follows. 

\subsection{Modulation/Verification Code Structure}

{\bf Code length.} The code consists of $n$ positions, $\alpha$ of which have energy, and the remaining $\beta = n-\alpha$ are empty, \ie absent of pulses (conceptually similar to OOK modulation, where $\alpha=\beta$). The code length affects the performance and security of the presented modulation technique. Larger $\alpha$ and $\beta$ values improve the security by reducing the probability of adversarial success in mounting undetectable distance-enlargement attack. However, increasing the code length reduces the frequency of conducting two-way ranging. Additionally, the Federal Communications Commission (FCC) imposes restrictions on the number of pulses with energy, effectively limiting $\alpha$ per unit of time. As such, $\beta$ could be independently increased to compensate for the loss of code length. Setting these parameters is discussed in Section~\ref{sec:eval}. 

%$(\Phi)$:
{\bf Pulse phase.} The sender uses a random-phase for the $\alpha$ pulses it transmits. Each phase is equally likely. The phase will be irrelevant for the receiver because ED receivers are agnostic to the phase, as explained in Section~\ref{sec:symbol_receiver}. The sender need not share this information with the receiver since the receiver measures the energy, not the polarity of the pulse.

%$(\phi)$:
{\bf Pulse permutation.} The sender and receiver secretly agree on a random permutation of the $n$ positions, obtained from a uniform distribution. Figure~\ref{fig:proposedsystemsmall} shows an example before and after the permutation. The verification code can thus be considered a sequence of $\{-1,0,1\}$ pulses, where $\{-1,1\}$ represent the phase, and $\{0\}$ pulse absence. %The transmitter and receiver share position of presence and absence of pulses, adversary is agnostic of this information however. 

{\bf Spacing between pulses.} The time between two consecutive pulses, $T_s$, is normally lower bounded by the delay spread of the channel. We submit that $T_s$ should be such that $T_s > 2d/\text{\bf c}$, where $d$ is the distance between the two devices. If the adversary replays the authentic signal delayed by more than the equivalent RTT, the attack will be detected by the mismatch between the measured RTT and the one equivalent to the committed distance. To avoid being detected, the adversary would thus replay its delayed version of a pulse within the $T_s$ time window. As such, authentic pulse $i$ will not overlap with the adversary's delayed version of pulse $i-1$, or any further adversary pulses $i-2$, $i-3$, \etc %That is, detection occurs without running the code detection technique presented herein. 

\newcommand{\bogus}{\cellcolor{white}}
\newcommand{\hpulse}{\cellcolor{gray!100}}
\newcommand{\lpulse}{\cellcolor{white}}

\begin{figure}
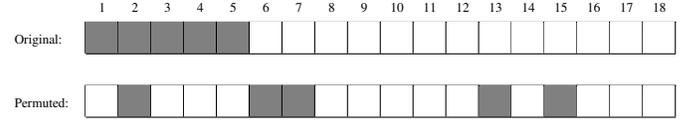

\centering
\begin{tabular}
{@{}l|c|c|c|c|c|c|c|c|c|c|c|c|c|c|c|c|c|c|l}
\multicolumn{1}{@{}c@{}}{}&
\multicolumn{1}{@{}c@{}}{\tiny 1}&
\multicolumn{1}{@{}c@{}}{\tiny 2}&
\multicolumn{1}{@{}c@{}}{\tiny 3}&
\multicolumn{1}{@{}c@{}}{\tiny 4}&
\multicolumn{1}{@{}c@{}}{\tiny 5}&
\multicolumn{1}{@{}c@{}}{\tiny 6}&
\multicolumn{1}{@{}c@{}}{\tiny 7}&
\multicolumn{1}{@{}c@{}}{\tiny 8}&
\multicolumn{1}{@{}c@{}}{\tiny 9}&
\multicolumn{1}{@{}c@{}}{\tiny 10}&
\multicolumn{1}{@{}c@{}}{\tiny 11}&
\multicolumn{1}{@{}c@{}}{\tiny 12}&
\multicolumn{1}{@{}c@{}}{\tiny 13}&
\multicolumn{1}{@{}c@{}}{\tiny 14}&
\multicolumn{1}{@{}c@{}}{\tiny 15}&
\multicolumn{1}{@{}c@{}}{\tiny 16}&
\multicolumn{1}{@{}c@{}}{\tiny 17}&
\multicolumn{1}{@{}c@{}}{\tiny 18}\\
\cline{2-19}

\tiny Original:
&
\hpulse &\hpulse &\hpulse &\hpulse &\hpulse &\lpulse &\lpulse &\lpulse &\lpulse &\lpulse &\bogus &\bogus &\bogus &\bogus &\bogus &\bogus &\bogus &\bogus \\
\cline{2-19}

\multicolumn{1}{@{}c@{}}{}\\
\cline{2-19}

\tiny Permuted:
&
\lpulse &\hpulse &\bogus &\bogus &\bogus &\hpulse &\hpulse &\bogus &\lpulse &\bogus &\bogus &\lpulse &\hpulse &\bogus &\hpulse &\lpulse &\lpulse &\bogus\\
\cline{2-19}

\end{tabular}
\caption{
An example verification code with a randomly-looking pulse reordering, where $\alpha=5$, $\beta=13$, and the code contains $n=\alpha+\beta=18$ pulses.
%-----------
Upon receiving the permuted code pulses as per the secret agreement between the sender and receiver, the receiver knows that Bin$_{\alpha}$ will contain the received energies at the positions (gray) \{2, 6, 7, 13, 15\}, which are the expected high-energy pulses. Bin$_{\beta}$ will contain the rest: \{1, 3, 4, 5, 8, 9, 10, 11, 12, 14, 16, 17, 18\}.
}
\label{fig:proposedsystemsmall}
\end{figure}

$~$

\begin{figure}
\centering
\includegraphics[scale=0.21]{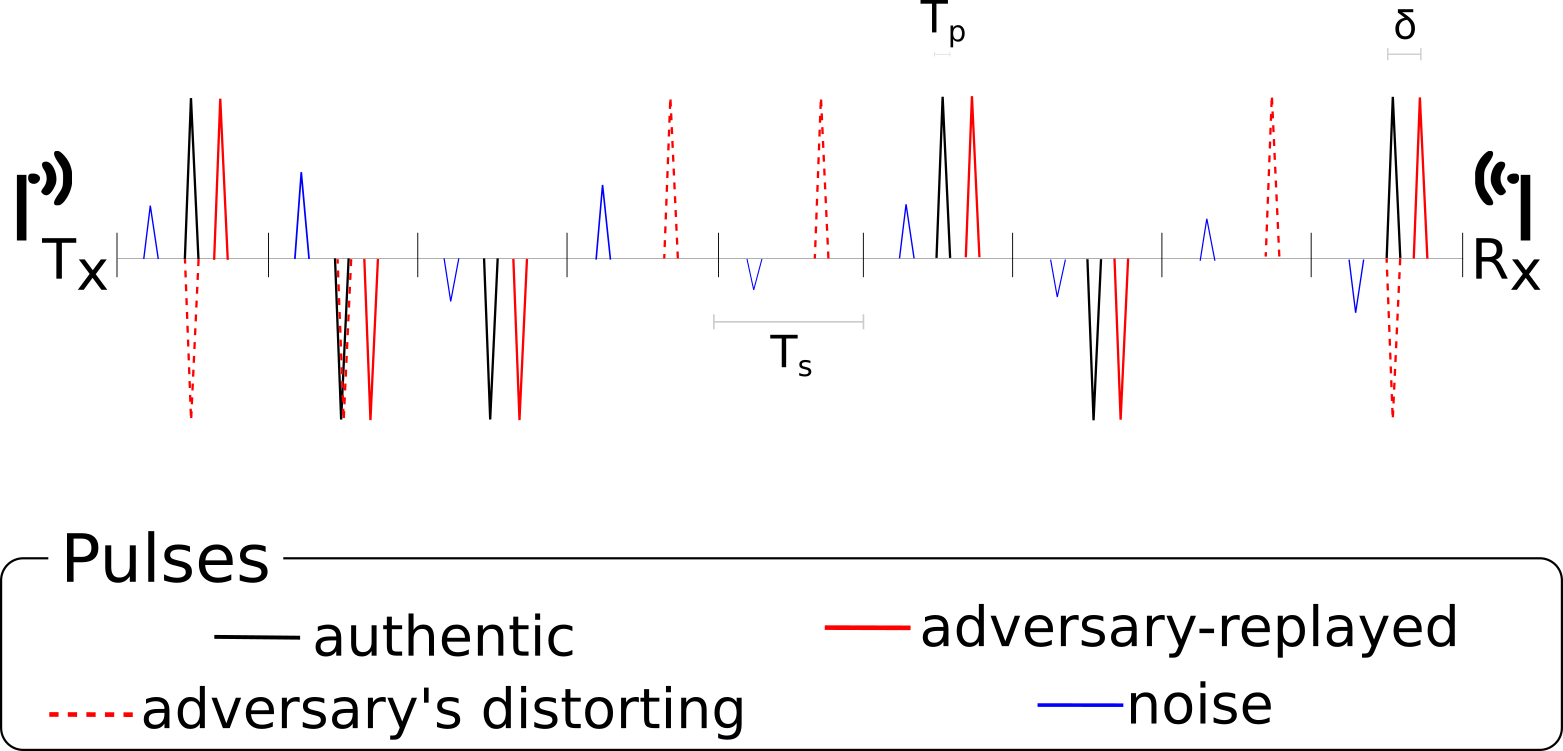}
\caption{An example verification code of $n$ slots (9 of which are shown),  the spacing  $T_s$ between consecutive pulses is  $1\mu s$ and pulse width $T_p$  is $2 ns$. An adversary transmits a pulse to distort the legitimate pulse (dashed red). The adversary also replays the authentic signal with the delay $\delta$ (solid red).
Best viewed in color.}
\label{fig:main}
\end{figure}

An example code structure, and adversarial attempts to corrupt and replay it, is shown in Fig.~\ref{fig:main}.

% $=3\cdot 10^8$ m/s

\subsection{Verification Code Identification}
\label{sec:vcodeident}

Upon receiving a transmission, the receiver starts processing the code associated with the highest preamble's peak. The code associated with a peak is the train of $T_s$-spaced pulses that start at a fixed time interval (\eg agreed upon between the sender and receiver) after the peak. This peak however may not be authentic, and could be the adversary's replayed version. 
The receiver thus backtracks at fixed time steps corresponding to the pulse width $T_p$ (\eg $2\ ns$), trying to identify if another version of the code (or a possible distorted imprint of it) was present in the transmission at an earlier time. The receiver does not need to backtrack further beyond some time $T_0$, knowing the maximum communication range. If the last distance verification occurred recently, the verified range could be used (in combination with the devices' upper bound motion speeds) to reduce the backtracking time.

Backtracking requires the receiver to record transmissions. If an earlier version of the code is found (and their difference exceeds the receiver's standard precision, \eg $\pm 10\ cm$ for DecaWave~\cite{deca}), it is used for ToF estimation. 

\begin{figure}
\includegraphics[width=1\linewidth]{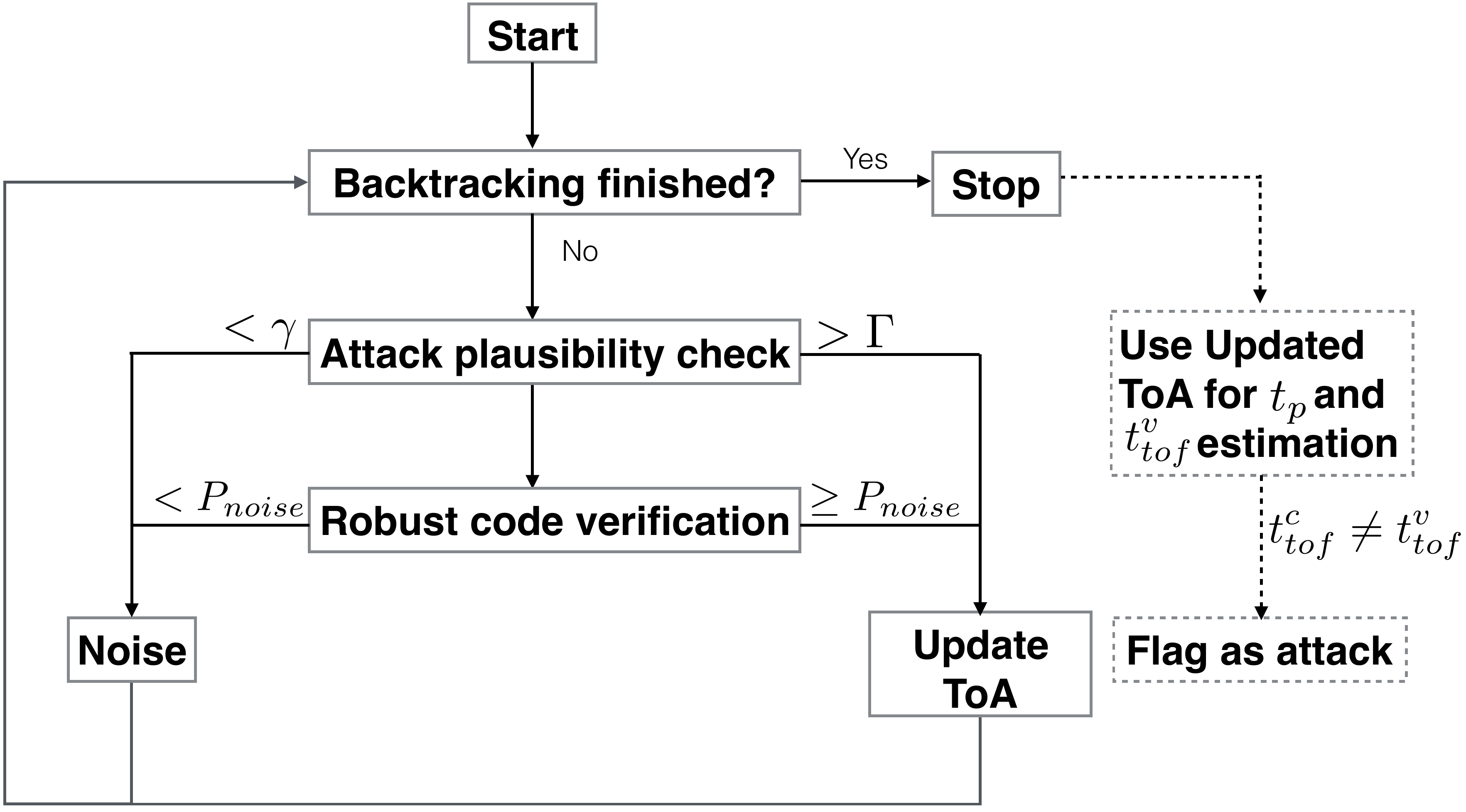}
\caption{The receiver backtracks to detect enlargement attacks. An event is flagged as an attack when the aggregate energy is higher than $\Gamma$ (\eg DoS, jamming), \ie the data looks more similar to a verification code than noise. The last flagged position is used for the ToF estimation.
 }
\label{fig:decisiontree}
\end{figure}

As shown in Fig.~\ref{fig:decisiontree}, the receiver performs \first{} check and \second{} to detect attacks until the maximum backtracking time is reached. For each code, the receiver does not look for an exact match of the transmitted pulses in their positions simply because that could be easily bypassed with minimal adversarial efforts (as explained in Section~\ref{sec:attackdesc}). Instead, the receiver proceeds as follows. Knowing the mapping of the pulse positions, the receiver distributes the received powers of each pulse among two bins, Bin$_\alpha$ and Bin$_\beta$. The former will have the values of the received power (\eg in Watts) of the energy-present pulse positions, the latter energy-absent positions (Fig.~\ref{fig:proposedsystemsmall}). 

{\bf \first{} check.} 
For each candidate verification code obtained during backtracking, the overall received signal power (the aggregate of Bin$_{\alpha}$ and Bin$_{\beta}$) is measured, and compared to a predefined threshold, $\gamma$. This threshold is based on the receiver's noise figure. If the aggregate exceeds $\gamma$, a potential verification code has been found. Otherwise it gets discarded as noise. The aggregate energy is then compared to another threshold, $\Gamma$. This is calculated based on the overall aggregate energy the receiver expects to receive based on the measured distance in the commitment phase, following the path loss model. Artificial distance enlargement caused by the adversary in the commitment phase lowers the receiver's calculated $\Gamma$ (because of the higher path loss), thus increases the likelihood of the actual received aggregate to exceed $\Gamma$. If the aggregate exceeds $\Gamma$, an adversary may possibly be injecting energy into the channel to distort the authentic code. If the verification code is neither discarded as noise ($< \gamma$) nor exceeds $\Gamma$, the receiver proceeds to the \second{} check.

{\bf \second{}.} 
Now the receiver checks the verification code content. If the receiver simply flags the presence of one or more pulses (above noise) in Bin$_\beta$ as an attack, false positives increase because such pulses could occur for many legitimate reasons (\eg noise spikes, reflections, interfering transmissions, antenna orientation, or multipath).\footnote{If the receiver instead interprets a pulse in Bin$_\beta$ as an indication that the code is not authentic and continues backtracking, it may very well skip the authentic code thus helping the adversary.} Instead, the receiver performs a sequence of binary hypothesis tests on random pulse samples. It tests  if the candidate code is more similar to an authentic code than noise. It chooses $r \le \alpha$ random pulses from the $\alpha$ in Bin$_\alpha$ (where $r$ is the number of pulses per symbol), aggregates their received powers and compares that to the aggregate of another $r$ pulses randomly chosen from the $\beta$ in Bin$_\beta$. If the aggregate of those selected from Bin$_\alpha$ is larger, the receiver identifies this as a candidate authentic code, and records its ToA. Finally, the distance is calculated based on the recorded ToA of the most recently received code, and a mismatch with the committed distance is flagged as an attack.

A candidate verification code could be again noise, which has slipped the \first{} check perhaps due to some sporadic noise spikes in the transmission. Noise has a probability of $\leq P_{\text{noise}}$ to satisfy the \second{} check, where $P_{\text{noise}}$ is derived as (\ref{eq:shorten}) in Section~\ref{sec:noisepassing}. As such, the receiver estimates the probability that the above condition is satisfied. This is done by repeating the random sampling $\upsilon$ times, and checking if the ratio of the number of times the condition is satisfied to $\upsilon$ exceeds $P_{\text{noise}}$. This would indicate the code is not noise, and is either authentic or adversary-replayed. 
Regardless, the receiver uses the ToA of the most recent code found.

\subsection{Setting the Energy Thresholds.}

{\bf Setting the upper-bound threshold, $\Gamma$.} To set $\Gamma$, the receiver relies on the committed (unverified) distance between itself and the sender. This dictates the path loss---the amount of power loss per pulse as pulses propagate the medium. Larger committed distance causes the receiver to expect less power, thus setting a lower $\Gamma$. Thus, by increasing the committed distance, the adversary helps divulge its malice.

The path loss function $f()$ for outdoor UWB LoS is~\cite{Channel_model_final_report}:
\begin{equation}
f(d) = PL_0  + 10\cdot n\cdot log\left(\frac{d}{d_0}\right)
\end{equation}
where $d$ is the distance in meters, and $PL_0$ is a constant representing the path loss at the reference distance $d_0$. For UWB LoS channel model, these constants are set to~\cite{Channel_model_final_report}:
\begin{equation}
\label{eq:pathloss}
f(d) = -46.3 - 20\ log(d) - log\left(\frac{6.5}{5}\right)
\end{equation}
This is calculated in the standard signal ratio unit, $dB$, where: 
\begin{equation}
\text{Power ratio (in } dB \text{)} = 10\ log\ (\text{ratio})
\end{equation}
The path loss function thus expresses the power loss as
\begin{equation}
f(d) = 10\ log\left(\frac{(\lambda_b)^2}{(\lambda_\text{sent})^2}\right)
\end{equation}
or
\begin{equation}
\label{eq:gammaflipped}
\frac{(\lambda_b)^2}{(\lambda_\text{sent})^2} = 10^{f(x)/10}
\end{equation}
where $(\lambda_b)^2$ is the pulse instantaneous power the receiver \emph{expects}, and $(\lambda_\text{sent})^2$ is that the sender has actually sent, \eg both in Watt. Knowing the constant pulse power of the sender, then the pulse power is expected to be received as:
\begin{equation}
(\lambda_b)^2 = (\lambda_\text{sent})^2 \ 10^{f(x)/10}
\end{equation}
The receiver then calculates $\Gamma$ as follows:
\begin{equation}
\label{eq:Gamma}
\begin{aligned}
\Gamma	&= \alpha\ (\lambda_b + N)^2 + \beta\ (N)^2
\end{aligned}
\end{equation}
%&= \alpha\ (\lambda_\text{sent}\ 10^{f(x)/10} + N)^2 + \beta\ (N)^2
where $d$ is the (unverified) distance in meters between the sender and receiver obtained at commit stage, either true or artificially enlarged in case of an attack. $N$ is an instantiation of zero-mean Gaussian noise at the receiver, \ie the noise present in the receiver's channel and cannot be removed \cite{Molisch_Wireless_Communication_book}.%(measured in the same unit as $\lambda_\text{sent}$).

There are other factors that contribute to the degradation of power. These factors could cause further power loss $E$, typically up to $E=-8\ dB$ more~\cite{Molisch_paper_channel_UWB,ChannelCondition_material}. If the receiver sets $\Gamma$ as that after the expected further degradation (\ie too small $\Gamma$ value), false positives may increase because such additional signal-degradation factors may or may not occur---if they do not, the receiver would then falsely assume such relatively ``too high'' aggregate energy is due to an attempted attack. Accordingly, the receiver sets $\Gamma$ based only on the (almost certain) path loss deterioration. Any further power loss would then be added benefit to the adversary, as it allows the adversary to inject more pulses into the channel to corrupt the authentic code without exceeding $\Gamma$. 

{\bf Setting the lower-bound threshold, $\gamma$.} If the aggregate energy is $< \gamma$, it would be either due to noise or a substantial deterioration of the authentic signal where no meaningful information could be recovered during the \second{}. Too high $\gamma$ leads to false negatives; too low triggers \second{} even for noise. For critical applications seeking to prevent false negatives, $\gamma$ could be set conservatively based on the receiver's noise variance $\sigma_N^2$:
\begin{equation}
\label{eq:gamma}
\begin{aligned}
\gamma	&= (\alpha + \beta)\cdot\sigma_N^2
\end{aligned}
\end{equation}

\subsection{Attack Resilience}
\label{sec:attkres}

Here we explain how UWB-ED resists standard enlargement attacks. More complex attacks are discussed in Section~\ref{sec:security}.

\subsubsection{Detecting Signal Replay} 

An adversary that simply replays authentic pulses does not win because the receiver backtracks to detect earlier copies of the code. UWB-ED provides resilience to benign signal distortion, \eg due to channel conditions or antenna orientation, because the receiver looks for similarities between the code and the received signal (versus exact data match), allowing for a higher bit error rate. In general, poor channel conditions (low SNR) can be compensated for by increasing the symbol length, $r$, minimizing the bit error rate.

\subsubsection{Complicating Signal Annihilation} %\vspace{-6pt}

The unpredictability of the pulse phase means an adversary must either wait to detect it and immediately inject the reciprocal pulse for annihilation, or inject a random-phased pulse hoping it is the reciprocal. The former is infeasible in practice for UWB (see Section~\ref{sec:threatmodel}). The latter results in amplifying or annihilating the authentic pulse, each with a 50\% chance. Amplification is unfortunate to the adversary, as the adversary now needs to compensate with an equivalent amplitude, $A$. Amplification doubles the amplitude. The estimated energy of the pulses will thus amount to $\sim A^2$, and the adversary-contributed amplification to $\sim (2A)^2$. 

Since the result is indeterministic for the adversary, it leads us to the next discussion: how successful would the adversary be in ``contaminating the evidence'' that an authentic verification code existed, and how much energy room does the adversary have to do that before exceeding $\Gamma$?

\subsubsection{Mitigating Evidence Contamination} %\vspace{-6pt}

\begin{figure}
\begin{tikzpicture}
\begin{axis}[
width=3.31in,
height=2.6in,
%axis lines=middle,
%ytick style={draw=none},
xmin=0,xmax=100,
%extra y ticks={0.000005708811108603221647105671,0.000001427202777150804988259944},
ymode=log,
xlabel={Distance ($m$)},
ylabel={Power loss ratio ($10^{f(x)/10}$)},
tick label style={font=\scriptsize},
label style={font=\scriptsize},
x label style={at={(0.5,0)}},
y label style={at={(0,0.5)}},
legend style={draw=none},
]
\addplot[thick,domain=1:100,samples=100] {10^((- 46.3 - 20*ln(x)/ln(10) - ln(6.5/5)/ln(10) )/10)};
\addlegendentry{\scriptsize Best receiver-expected signal}

\addplot[dashed,domain=1:100,samples=100] {10^((- 46.3 - 20*ln(x)/ln(10) - ln(6.5/5)/ln(10) - 5)/10)};
\addlegendentry{\scriptsize $E=-5\ dB$}

\addplot[,domain=1:100,samples=100] {10^((- 46.3 - 20*ln(x)/ln(10) - ln(6.5/5)/ln(10) - 10)/10)};
\addlegendentry{\scriptsize $E=-10\ dB$ (worst)}

%\addplot[dotted,domain=1:100,samples=100] {10^((- 46.3 - 20*ln(x)/ln(10) - ln(6.5/5)/ln(10) - 20)/10)};
%\addlegendentry{\scriptsize $E=-20\ dB$ (worst)}

%\addplot[red,domain=1:300,samples=200] {- 46.3 - 20*ln(x)/ln(10) - ln(6.5/5)/ln(10)};
%\addlegendentry{\scriptsize In $dB$}

\draw[dashed,red,line width=0mm] (axis cs:30.22,10^-15) -- (axis cs:30.22,0.000000025004417674767125972474);
\draw[,red,line width=0mm] (axis cs:0,0.000000025004417674767125972474) -- (axis cs:30,0.000000025004417674767125972474);

%\draw[dashed,red!60,line width=0mm] (axis cs:47.79,10^-15) -- (axis cs:47.79,0.000000009998425241637153044683);
%\draw[,red!60,line width=0mm] (axis cs:0,0.000000009998425241637153044683) -- (axis cs:47.79,0.000000009998425241637153044683);

\draw[dashed,green,line width=0mm] (axis cs:15.11,10^-15) -- (axis cs:15.11,0.000000010001767069906852705095);
\draw[,green,line width=0mm] (axis cs:0,0.000000010001767069906852705095) -- (axis cs:15.11,0.000000010001767069906852705095);

\node[anchor=west] (source) at (axis cs:0,10^-9.2){\tiny Actually-received signal};
\node (destination) at (axis cs:15.11,0.000000010001767069906852705095){};
\draw[->,>=stealth](source) to [out=155,in=200] (destination);
%\draw[->](source)--(destination);

\node[anchor=west] (source) at (axis cs:30,10^-6.9){\tiny Receiver's threshold per pulse};% (at $D_2=14.89$)};
\node (destination) at (axis cs:30.22,0.000000025004417674767125972474){};
\draw[->,>=stealth](source) to [out=180,in=100] (destination);
%\draw[->](source)--(destination);

%\node[anchor=west] (source) at (axis cs:40,10^-7.5){\tiny Receiver's threshold (at $D_2=32.68$)};
%\node (destination) at (axis cs:47.79,0.000000009998425241637153044683){};
%\draw[->,>=stealth](source) to [out=-90,in=5] (destination);

\end{axis}
\end{tikzpicture}
\caption{The best expected signal power as calculated by the receiver using the path loss function in (\ref{eq:pathloss}), the signal at $E=-5\ db$ of further power loss, and at $E=-10\ db$ (worst expected). If the distance is $D_1 = 15.11\ m$ (green line), and the adversary doubles it, \ie by adding $D_2=15.11\ m$ to make it $D_1+D_2 = 30.22\ m$ (red line), the receiver will set the threshold following the fake distance, at $10^{f(D_1+D_2)/10} = 10^{-7.6}$. The adversary's room is the difference between the red and green lines on the $y$-axis. At $D_2 = 32.68\ m$, the adversary has no room. Best viewed in color.}
\label{fig:pathloss}
\end{figure}
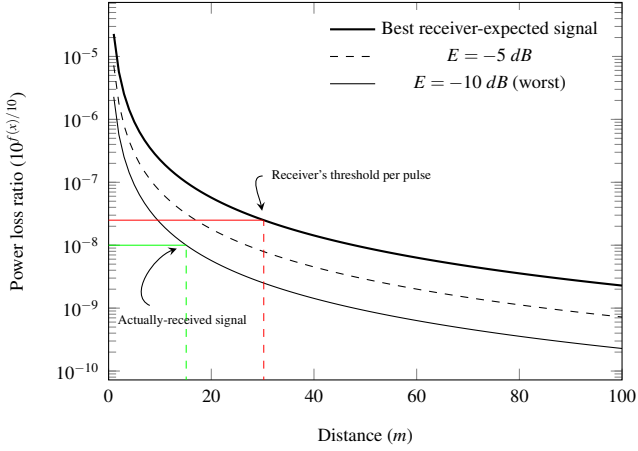
%$D_2 = 32.68\ m$ to make it $D_1+D_2 = 47.79\ m$ (red line)
To hide the authentic code, the adversary tries to inject energy into the channel, hoping it annihilates as many of Bin$_\alpha$ pulses as possible. We thus calculate the room available to the adversary here, and use that to derive the probability of adversarial success in distance enlargement in Section~\ref{sec:eval}.

Figure~\ref{fig:pathloss} shows the path loss function in (\ref{eq:gammaflipped}) as used by the receiver to detect the threshold $\Gamma$, as well as the worst receiver-expected signal after additional deterioration. The receiver sets the threshold based on the best expected signal. The room available for the adversary to add energy depends on the actual signal received. The most favorable situation to the adversary is when the received signal power is the worst (lowest $E$), which allows the adversary to inject pulses without exceeding $\Gamma$. For example, in Fig.~\ref{fig:pathloss}, if the actual distance between the sender and receiver is $D_1 = 15.11\ m$ (green line), and the adversary is trying to add $D_2 = 32.68\ m$ to make the distance $D_1+D_2 = 47.79\ m$ (red line), the receiver will set $\Gamma$ using the fake distance, $D_1+D_2$. At such a relatively large added distance, $D_2$, the received pulse power is unlikely to fall below $f(D_1)+E = 10^{-8}(\lambda_\text{sent})^2$ at, \eg $E=-10\ dB$. The room available to the adversary to inject energy becomes too small, significantly reducing its chances of success.

\begin{figure}
\centering
%\subfloat[$\beta=100$]{
%\label{}
\begin{tikzpicture}
\begin{axis}[
width=3.31in,
height=\hghtthird,
xlabel={Adversary-added distance ratio ($D_2/D_1$)},
ylabel={Adversary room ($\zeta$)},
xmin=0, xmax=3,
ymin=0, ymax=12,
%ytick ={0,1},
%xtick ={0,100,200},
%extra x ticks={1.5},
extra x tick style={grid=major},
%extra y ticks={0.35,0.54},
extra y tick style={grid=major,yticklabels={,,}},
%minor x tick num=1,
grid=minor,
%xmode=log,
tick label style={font=\scriptsize},
x label style={font=\scriptsize,at={(0.5,0.07)}},
y label style={font=\scriptsize,at={(0.07,0.5)}},
legend columns=1,
legend style={draw=none,at={(0.2,1)},anchor=north,}
]
%\addplot[dotted] table[col sep=comma]{csv/zetaE-20.csv};
%\addlegendentry{\scriptsize $E=-20\ dB$}
\addplot[] table[col sep=comma]{csv/zetaE-10.csv};
\addlegendentry{\scriptsize $E=-10\ dB$}
\addplot[dashed] table[col sep=comma]{csv/zetaE-5.csv};
\addlegendentry{\scriptsize $E=-5\ dB$}
\end{axis}
\end{tikzpicture}
\caption{Adversary's room to add energy, $\zeta$ in (\ref{eq:zeta}), against the ratio of the adversary-added to true distance ($D_2/D_1$); $E$ represents additional signal degradation beyond path loss.}
\label{chart:zeta}
\end{figure}
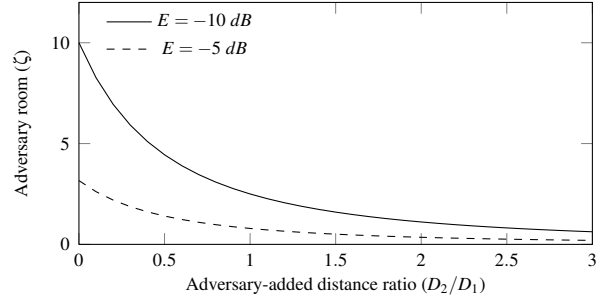

The room-per-pulse, $R$, available to the adversary to enlarge the distance thus lies in-between the received signal and $\Gamma$, and is calculated in $dB$ as:
\begin{equation}
\label{eq:rperpulse}
R = f(D_1+D_2) - (f(D_1)+E)
\end{equation}
where $E$ represents other channel degrading factors, and the distances $D_1$ and $D_2$ (in meters) are respectively the true distance between both devices, and the extra distance the adversary intends to add. This room is thus expressed as:
\begin{equation}
\label{eq:zeta}
\zeta=10^{R/10}
\end{equation}
Figure~\ref{chart:zeta} plots $\zeta$ at various distance ratios $D_2/D_1$.

Recall that the adversary may succeed to annihilate some of the pulses falling in Bin$_\alpha$. But since Bin$_\beta$ in the authentic code have nothing but noise, adding pulses into those will result in an increase in the overall aggregate energy. As such, this available energy room in (\ref{eq:rperpulse}) by itself does not give a perfect indication to the adversary's chances of success.

\subsection{A Numerical Example}

\begin{figure*}
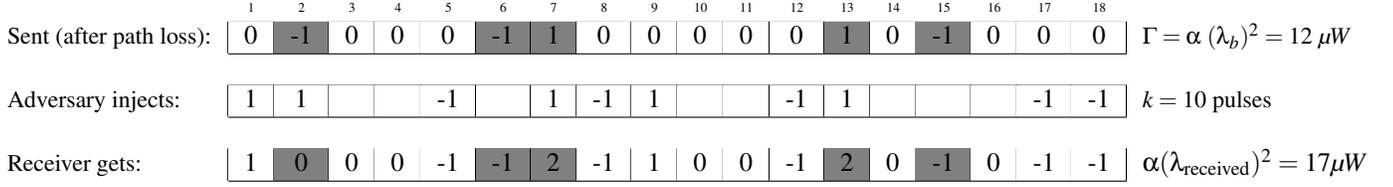

\centering
\begin{tabular}
{@{}l|c|c|c|c|c|c|c|c|c|c|c|c|c|c|c|c|c|c|l}
\multicolumn{1}{@{}c@{}}{}&
\multicolumn{1}{@{}c@{}}{\tiny 1}&
\multicolumn{1}{@{}c@{}}{\tiny 2}&
\multicolumn{1}{@{}c@{}}{\tiny 3}&
\multicolumn{1}{@{}c@{}}{\tiny 4}&
\multicolumn{1}{@{}c@{}}{\tiny 5}&
\multicolumn{1}{@{}c@{}}{\tiny 6}&
\multicolumn{1}{@{}c@{}}{\tiny 7}&
\multicolumn{1}{@{}c@{}}{\tiny 8}&
\multicolumn{1}{@{}c@{}}{\tiny 9}&
\multicolumn{1}{@{}c@{}}{\tiny 10}&
\multicolumn{1}{@{}c@{}}{\tiny 11}&
\multicolumn{1}{@{}c@{}}{\tiny 12}&
\multicolumn{1}{@{}c@{}}{\tiny 13}&
\multicolumn{1}{@{}c@{}}{\tiny 14}&
\multicolumn{1}{@{}c@{}}{\tiny 15}&
\multicolumn{1}{@{}c@{}}{\tiny 16}&
\multicolumn{1}{@{}c@{}}{\tiny 17}&
\multicolumn{1}{@{}c@{}}{\tiny 18}\\
\cline{2-19}

\small Sent (after path loss):&
\lpulse 0&\hpulse -1&\bogus 0&\bogus 0&\bogus 0&\hpulse -1&\hpulse 1&\bogus 0&\lpulse 0&\bogus 0&\bogus 0&\lpulse 0&\hpulse 1&\bogus 0&\hpulse -1&\lpulse 0&\lpulse 0&\bogus 0
&\small $\Gamma	= \alpha\ (\lambda_b)^2 = 12\ \mu W$\\
\cline{2-19}

\multicolumn{1}{@{}c@{}}{}\\\cline{2-19}

\small Adversary injects:&
\bogus 1&\bogus 1&\bogus &\bogus &\bogus -1&\bogus &\bogus 1&\bogus -1&\bogus 1&\bogus &\bogus &\bogus -1&\bogus 1&\bogus &\bogus &\bogus &\bogus -1&\bogus -1
&\small $k=10$ pulses\\
\cline{2-19}

\multicolumn{1}{@{}c@{}}{}\\\cline{2-19}

\small Receiver gets:&
\lpulse 1&\hpulse 0&\bogus 0&\bogus 0&\bogus -1&\hpulse -1&\hpulse 2&\bogus -1&\lpulse 1&\bogus 0&\bogus 0&\lpulse -1&\hpulse 2&\bogus 0&\hpulse -1&\lpulse 0&\lpulse -1&\bogus -1
&$\alpha (\lambda_\text{received})^2 = 17 \mu W$\\
\cline{2-19}

\end{tabular}
\caption{An example of the random-phased Bin$_\alpha$ pulses (dark gray) reordered following the permutation in Fig.~\ref{fig:proposedsystemsmall}.
%-----------
After the adversary injects $k=10$ random-phased pulses at random positions, the receiver will get the summation at each pulse position.
%-----------
%If the receiver chooses $r=2$ pulses at random from Bin$_\beta$, there is a probability of 0.12 (in this particular example) that their aggregate will exceed the aggregate of any $2$ chosen from Bin$_\alpha$. This is the probability of the attack succeeding (provided the overall aggregate did not exceed the receiver's threshold) for this particular distribution of sender and adversary pulses. See equation (\ref{eq:probatksuccesspart1}) in Section~\ref{sec:probfirstcheck} for a general derivation of this probability given $n$, $\alpha$, $k$, and $r$.
%-----------
%Note: the aggregate absolute energy as sent is $5 mW$, and of the received is $13 mW$, which is $13/5 = 2.6$ times what the receiver expected to receive, or $10\ log(2.6) = 4.1$ dB more.
}
\label{fig:proposedsystem}
\end{figure*}
%At r=1: probability of success is 4/25 = 0.160000
%At r=2: probability of success is 12/100 = 0.120000
%At r=3: probability of success is 4/100 = 0.040000
%At r=4: probability of success is 0/25 = 0.000000
%At r=5: probability of success is 0/1 = 0.000000

Figure~\ref{fig:proposedsystem} shows an example verification code, expanded from Fig.~\ref{fig:proposedsystemsmall}, where the adversary injects $k=10$ random-phased pulses. For simplicity, the figure assumes $N=0$. If the distance between the sender and receiver is $D_1=4\ m$, and the adversary is trying to enlarge it by $D_2=4.5\ m$ to make it $D_1+D_2=8.5\ m$, and assuming $(\lambda_\text{sent})^2 = 7.6\ \mu W$, then the receiver expects a best case received power of:
\begin{equation}
\label{eq:lambdab}
\begin{aligned}
(\lambda_b)^2	&= (\lambda_\text{sent})^2\ 10^{f(D_1+D_2)/10} \\
&=  7.67\times 10^{f(8.5)/10} = 2.4\ \mu W\\
\end{aligned}
\end{equation}
From (\ref{eq:gamma}) at $N=0$ and $\alpha=5$ (as in Fig.~\ref{fig:proposedsystem}), it then calculates the threshold as:
\begin{equation}
\Gamma	= \alpha\ (\lambda_b)^2 = 12\ \mu W
\end{equation}

At $E=-10\ dB$, the actual signals are received as:
\begin{equation}
\label{eq:lambdaw}
(\lambda_w)^2	= (\lambda_\text{sent})^2\ 10^{(f(D_1)+E)/10}\approx  1\ \mu W
\end{equation}

Now assuming the adversary is $D_3 = 6\ m$ away from the receiver, and uses a random-phased pulse with transmission power of  $(\lambda_\text{sent}^\text{adversary})^2 = 15.77\ \mu W$. At $E=-10\ dB$, the receiver would receive the adversary's signals as:
\begin{equation}
\label{eq:advepower}
(\lambda')^2	= (\lambda_\text{sent}^\text{adversary})^2\ 10^{(f(D_3)+E)/10}\approx  1\ \mu W
\end{equation}

%As for the sender's pulses, the receiver would actually (\ie based on the actual distance) receive them as:

So in the best case for the adversary, where the signal is highly deteriorated, the adversary would then have a per-pulse room of $R = 3.45\ dB$ to add energy, which amounts to 7 $\mu W$ more, \ie up to $\Gamma=12\mu W$. In Fig.~\ref{fig:proposedsystem}, after the adversary injects its $k=10$ pulses at the example random positions and with the random phases shown, it results in annihilating a single pulse (at position 2), amplifying two pulses (at positions 7 and 13), and adding seven more 1 $\mu W$ pulses for an increase of the overall aggregate to be 17 $\mu W$. This exceeds $\Gamma=12\ \mu W$, and this attack would thus be detected.% even if the \second{} check fails.

\section{Evaluation}
\label{sec:eval}

We evaluate UWB-ED by deriving the probability of success for an adversary enlarging the distance. We also validate that model using simulations in Section~\ref{sec:validate}.

\newcommand{\chse}[2]{{#1\choose #2}}

\subsection{Probability of a Successful Attack}
\label{sec:probsuccess}

The adversary hides the authentic code by having the aggregate of the $r$ pulses that the receiver chooses from Bin$_\beta$ exceed Bin$_\alpha$. The adversary must also avoid injecting too much energy to not exceed $\Gamma$. Not knowing which pulse belongs to which bin, the adversary injects $k$ pulses at random positions thus affecting $k$ of the $n$ pulses in the code.

To that end, the probability of mounting a successful attack, $P_{sa}$, is the intersection of the probability of two events (the checks in Fig.~\ref{fig:decisiontree}): the aggregate of the energy pulses chosen from Bin$_\beta$ ($b\beta$) exceeds that of Bin$_\alpha$ ($b\alpha$), and the added energy is $\leq \Gamma$:
\begin{equation}
\label{eq:finalprob}
P_{\text{sa}}(\alpha, \beta, r, \Gamma, k) = P_{b\beta>b\alpha}(\alpha, \beta, r, k)\cap P_{\leq \Gamma}(\alpha, \beta, k)
\end{equation}

\subsubsection{Probability of successfully evading the \second{} check ($P_{b\beta>b\alpha}$)}
\label{sec:probfirstcheck}

To evade this, the adversary must have an energy aggregated from Bin$_\beta$ exceed Bin$_\alpha$. When the adversary injects $k$ pulses into the channel, $x$ will fall into Bin$_\alpha$, and the remaining $k-x$ into Bin$_\beta$. $P_{b\beta>b\alpha}$ is then the probability of this distribution occurring multiplied by the probability of the attack succeeding under this distribution, for all possible such distributions $0\leq x\leq \alpha$ and $0\leq k-x\leq \beta$. To calculate the probability of the distribution occurring, consider the general case of a bucket containing two types of objects (\eg colored pearls): $I$ of the first type, and $J$ of the second. If $\psi$ objects are selected at random, the probability that $i$ and $j$ of the $\psi$ are respectively of the first and second type ($i+j=\psi$) is:
\begin{equation}
\label{eq:main}
\frac{\chse{I}{i}\ \chse{J}{j}}{\chse{I+J}{i+j}}
\end{equation}
where $\chse{n}{r}$ denotes $n$ choose $r$ and is given by:
\begin{equation*}
\chse{n}{r} = 
\begin{cases}
\displaystyle \frac{n!}{r!(n-r)!}	,	&0\leq r\leq n\\[4pt]
0									,	&\text{otherwise}
\end{cases}
\end{equation*}

Similarly, the probability that $x$ and $k-x$ of the adversary's $k$ pulses respectively affect the $\alpha$ in Bin$_\alpha$ and $\beta$ in Bin$_\beta$ is:
\begin{equation*}
\frac{\chse{\alpha}{x} \  \chse{\beta}{k-x}}{\chse{\alpha+\beta}{k}}
\end{equation*}
For all possible such distributions, we have:
\begin{equation}
\label{eq:probatksuccesspart1}
P_{b\beta>b\alpha}(\alpha, \beta, r, k) = \sum_{x=0}^\alpha \left(p_{\alpha, \beta, r, k}(x) \cdot \frac{\chse{\alpha}{x} \  \chse{\beta}{k-x}}{\chse{\alpha+\beta}{k}}\right)
\end{equation}
where $p_{\alpha, \beta, r, k}(x)$ is the probability $b\beta>b\alpha$ given the adversary affected $x$ and $k-x$ pulses in Bin$_\alpha$ and Bin$_\beta$ respectively.

To derive $p_{\alpha, \beta, r, k}(x)$, we assume for simplicity a unity power-per pulse, \ie the sender's and the adversary's pulses reach the receiver after path loss and other factors at a constant energy of $\pm 1 \mu W$.\footnote{Analogous analysis applies for non-constant energy.} This is similar to the example given in Fig.~\ref{fig:proposedsystem}. Every adversary-added pulse in Bin$_\beta$ will result in a 1 $\mu W$ of added energy from the receiver's point of view since the receiver's aggregation is agnostic to a pulse's phase. For Bin$_\alpha$, after the adversary affects $x$ pulses, some will be annihilated while others will be amplified. From the receiver's point of view, after the adversary's pulses are injected, Bin$_\alpha$ will have a mix of $2^2 = 4 \mu W$ and 0 $\mu W$ (adversary-affected) pulses, as well as the original 1 $\mu W$ unaffected pulses.

%Intuitively, if $g$ pulses result in a signal annihilation (\ie the original pulse plus that of the adversary gives zero), 
More $0\ \mu W$ (annihilated) pulses in Bin$_\alpha$ raises the chances that $b\beta>b\alpha$, which is in the adversary's favor. Since every affected pulse in Bin$_\alpha$ will either result in a $0\ \mu W$ or a $4\ \mu W$ pulse, there are $2^{x}$ possible outcomes. Of those, there are $\chse{x}{g}$ ways that $g$ $0\ \mu W$ pulses will occur. The probability that the $x$ adversary-injected pulses that fell in Bin$_\alpha$ result in a annihilation of $g$ pulses is thus $\chse{x}{g}/(2^{x})$.
%The probability of $g$ zeros occurring is thus the ratio of the number of ways of choosing $g$ from $x_1$ to the total number of possibilities~$2^{x_1}$.
For all possible numbers of annihilated pulses $0\leq g\leq x$, the adversarial success probability in the event that $x$ fell in Bin$_\alpha$ is:
\begin{equation}
\label{eq:fine}
p_{\alpha, \beta, r, k}(x) = \sum_{g=0}^{x} \left(p_{\alpha, \beta, r, k, x}(g) \cdot \frac{\chse{x}{g}}{2^{x}}\right)
\end{equation}
where $p_{\alpha, \beta, r, k, x}(g)$ is the probability $b\beta>b\alpha$ given $g$ annihilated pulses in Bin$_\alpha$.

When Bin$_\alpha$ has $g$ annihilated (0 $\mu W$), $x-g$ amplified (4 $\mu W$), and $\alpha-x$ unaffected pulses (1 $\mu W$), the probability of $b\beta>b\alpha$ in the event $x$ fell in Bin$_\alpha$, and $g$ of the $x$ pulses were annihilated is the probability that an aggregate of $m-1$ is chosen from Bin$_\alpha$ and an aggregate of $\geq m$ is chosen from Bin$_\beta$. For each possible $0\leq y_1,y_2\leq r$, we have:\\[1em]\noindent % (for all $0<m\leq r$) 
\begin{math}
p_{\alpha, \beta, r, k, x}(g) = 
\end{math}
\begin{equation}
\label{eq:finest}
\sum_{y_1=0}^r \sum_{y_2=0}^r \left(\frac{\chse{g}{y_1} \  \chse{x-g}{y_2} \  \chse{\alpha-x}{r-y_1-y_2}}{\chse{\alpha}{r}} \cdot \sum_{i=m}^r \frac{\chse{k-x}{i} \  \chse{\beta-(k-x)}{r-i}}{\chse{\beta}{r}}\right)
\end{equation}
where $m$ is:
%	&= 4y_2+r-(y_1+y_2)+1\\
\begin{equation}
\begin{aligned}
m	&= 0^2 \times y_1 + 2^2 \times y_2+1^2 \times (r-(y_1+y_2))+1\\
	&= r-y_1+3y_2+1
\end{aligned}
\end{equation}
At $r=\alpha$ (\ie selecting all Bin$_\alpha$ pulses) and $\alpha\leq\beta$, we get:
\begin{equation}
p_{\alpha, \beta, r, k, x}(g) = \sum_{i=m'}^r \frac{\chse{k-x}{i} \  \chse{\beta-(k-x)}{r-i}}{\chse{\beta}{r}}
\end{equation}
where $m'$ is:
\begin{equation}
\begin{aligned}
m'	&= 2^2 \times (x-g) + 1^2 \times (\alpha-x) + 1\\
	&= 4 (x-g) + (\alpha-x) + 1
\end{aligned}
\end{equation}

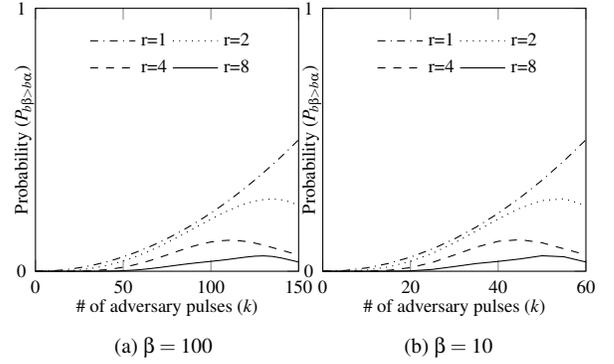
\begin{figure}
\centering
\subfloat[$\beta=100$]{
\label{awelwa7ed}
\begin{tikzpicture}
\begin{axis}[
width=\wdth,
height=\hght,
xlabel={\# of adversary pulses ($k$)},
ylabel={Probability ($P_{b\beta>b\alpha}$)},
xmin=0, xmax=150,
ymin=0, ymax=1,
ytick ={0,1},
%xtick ={0,100,200},
%extra x ticks={1.5},
extra x tick style={grid=major},
%extra y ticks={0.35,0.54},
extra y tick style={grid=major,yticklabels={,,}},
%minor x tick num=1,
%grid=minor,
tick label style={font=\scriptsize},
x label style={font=\scriptsize,at={(0.5,0.08)}},
y label style={font=\scriptsize,at={(0.31,0.5)}},
legend columns=2,
legend style={draw=none,at={(0.52,0.95)},anchor=north,}
]
\addplot[dashdotted] table[col sep=comma]{csv/prob/first/a50b100r1.csv};
\addlegendentry{\scriptsize r=1}
\addplot[dotted] table[col sep=comma]{csv/prob/first/a50b100r2.csv};
\addlegendentry{\scriptsize r=2}
\addplot[dashed] table[col sep=comma]{csv/prob/first/a50b100r4.csv};
\addlegendentry{\scriptsize r=4}
\addplot[] table[col sep=comma]{csv/prob/first/a50b100r8.csv};
\addlegendentry{\scriptsize r=8}
\end{axis}
\end{tikzpicture}
}
\hspace{\spcbtwn}
\subfloat[$\beta=10$]{
\label{tanywa7ed}
\begin{tikzpicture}
\begin{axis}[
width=\wdth,
height=\hght,
xlabel={\# of adversary pulses ($k$)},
ylabel={Probability ($P_{b\beta>b\alpha}$)},
xmin=0, xmax=60,
ymin=0, ymax=1,
ytick ={0,1},
%xtick ={0,50,100},
%extra x ticks={1.5},
extra x tick style={grid=major},
%extra y ticks={0.35,0.54},
extra y tick style={grid=major,yticklabels={,,}},
%minor x tick num=1,
%grid=minor,
tick label style={font=\scriptsize},
x label style={font=\scriptsize,at={(0.5,0.08)}},
y label style={font=\scriptsize,at={(0.31,0.5)}},
legend columns=2,
legend style={draw=none,at={(0.52,0.95)},anchor=north,}
]
\addplot[dashdotted] table[col sep=comma]{csv/prob/first/a50b10r1.csv};
\addlegendentry{\scriptsize r=1}
\addplot[dotted] table[col sep=comma]{csv/prob/first/a50b10r2.csv};
\addlegendentry{\scriptsize r=2}
\addplot[dashed] table[col sep=comma]{csv/prob/first/a50b10r4.csv};
\addlegendentry{\scriptsize r=4}
\addplot[] table[col sep=comma]{csv/prob/first/a50b10r8.csv};
\addlegendentry{\scriptsize r=8}
\end{axis}
\end{tikzpicture}
}
\caption{Probability that the \second{} check fails to detect the adversary's attack, plotted using (\ref{eq:probatksuccesspart1}) in Section~\ref{sec:probfirstcheck}, at $\alpha=50$ and $0\leq k\leq \alpha+\beta$.}
\label{chart:firstcheck}
\end{figure}

Figure~\ref{chart:firstcheck} plots $P_{b\beta>b\alpha}$, where $\alpha=50$. From these results, increasing $\beta$ is not necessarily effective for the \second{} check to detect attacks, since the adversary maintains its success probability by increasing $k$ proportionally; there is a visually similar pattern of adversarial success probability in both Fig.~\ref{awelwa7ed} and \ref{tanywa7ed}. As such, the advantage of the empty pulses in Bin$_\beta$ does not quite manifest in the \second{} check, rather the \first{} check.

Another observation is that higher $r$ lowers the adversary's success probability. For example at $\beta=100$ (Fig.~\ref{awelwa7ed}), the adversary has a 27\% chance at $r=2$ (which occurs at $k=135$), versus 5.85\% at $r=8$ (at $k=130$). In Section~\ref{sec:selectingrsecurity}, we show that at $r=\alpha$, we get the optimal security results.

\subsubsection{Final Probability of Adversary's Success}
\label{sec:finalprob}

In (\ref{eq:finalprob}), the event that the aggregate energy after the adversary's pulses is $\leq \Gamma$ and the event that $b\beta>b\alpha$ are dependent, and thus their intersection is not their product. Recall that in (\ref{eq:fine}), $g$ is the number of annihilated pulses, $x-g$ is the number of amplified pulses in Bin$_\alpha$, and $k-x$ is the number of added pulses in Bin$_\beta$. The aggregate-energy does not exceed $\Gamma$ when the adversary's pulses satisfy the inequality:
\begin{equation}
\label{eq:toadd}
\begin{split}
&(k-x)\ (\lambda'+N)^2 + (x-g)\ (\lambda'+\lambda_w+N)^2 +\\
&(\alpha-x)\ (\lambda_w+N)^2 + (\beta-(k-x)+g)\ (N)^2 \leq \Gamma
\end{split}
\end{equation}
where $\lambda'$ is defined as in (\ref{eq:advepower}), and $\Gamma$ in (\ref{eq:gamma}). 

If the adversary uses a variable pulse power randomly chosen from a distribution with a mean much different from $\lambda_w$, authentic pulses colliding with their reciprocal will not be fully annihilated. The adversary thus sets its power such that its mean at the receiver matches the sender, \ie $\overline{(\lambda')^2} = (\lambda_w)^2$. Assuming $(\lambda_w)^2 = (\lambda')^2$ in (\ref{eq:toadd}), we get:
\begin{equation}
\label{indeq:towards}
k+2x-4d+\alpha	\leq \frac{\alpha\ \lambda_b^2-\epsilon}{\lambda_\text{w}^2}
\end{equation}
where $\epsilon$ is a representation of noise, and evaluates to:
\begin{equation*}
\epsilon = N\ (\lambda_w\ (2k+2\alpha-4g)-\lambda_b(2\alpha))
\end{equation*}
As $\epsilon\rightarrow 0$, (\ref{indeq:towards}) becomes:
\begin{equation}
k+2x-4d	\leq \alpha\ \left(\frac{\lambda_b^2}{\lambda_\text{w}^2}-1\right)
\end{equation}
From (\ref{eq:lambdab}) and (\ref{eq:lambdaw}), we have:
\begin{equation}
\begin{aligned}
\frac{\lambda_b^2}{\lambda_\text{w}^2}	&= \frac{(\lambda_\text{sent})^2\ 10^{f(D_1+D_2)/10}}{(\lambda_\text{sent})^2\ 10^{(f(D_1)+E))/10}}\\
										&=10^{(f(D_1+D_2)-(f(D_1)+E))/10}\\
										&=\zeta\\
\end{aligned}
\end{equation}
where $\zeta$, from (\ref{eq:zeta}), represents the room-per-pulse available to the adversary to add energy into the channel.

We now calculate $p_{\alpha, \beta, r, k}(x,\Gamma)$, similar to (\ref{eq:fine}) as:
\begin{equation}
\label{eq:newpar}
p_{\alpha, \beta, r, k}(x,\Gamma) = \sum_{g=0}^{x} \left(p_{\alpha, \beta, r, k, x,\Gamma}(g) \cdot \frac{\chse{x}{g}}{2^{x}}\right)
\end{equation}
such that
\begin{equation}
p_{\alpha, \beta, r, k, x,\Gamma}(g)= 
\begin{cases}
\displaystyle p_{\alpha, \beta, r, k, x}(g)	,	&k+2x-4d\leq \alpha (\zeta-1)\\[4pt]
0												,	&\text{otherwise}
\end{cases}
\end{equation}
%and $p_{\alpha, \beta, r, k, x}(g)$ is defined as in (\ref{eq:finest}). Expressed in $\Gamma$, $\zeta$ evaluates to:
%\begin{equation}
%\zeta = \frac{1}{\lambda_w}\left(\frac{\Gamma-N\ (\alpha+\beta)}{\alpha}-N\right)
%\end{equation}

Using (\ref{eq:newpar}), the final adversarial success probability is:
\begin{equation}
\label{eq:finalprobcalc}
P_{\text{sa}}(\alpha, \beta, r, \Gamma, k) = \sum_{x=0}^\alpha \left(p_{\alpha, \beta, r, k}(x,\Gamma) \cdot \frac{\chse{\alpha}{x} \  \chse{\beta}{k-x}}{\chse{\alpha+\beta}{k}}\right)
\end{equation}

Figures~\ref{khameswa7ed} and \ref{sadeswa7ed} plot $P_{\text{sa}}$ in (\ref{eq:finalprobcalc}). At $\zeta=20$, $\Gamma$ is too high to reduce $P_{\text{sa}}$, but the \second{} check enables the receiver to limit it to $P_{\text{sa}}< 0.16\times 10^{-3}$. 
At $\zeta=10$, $P_{\text{sa}}$ stops growing beyond $0.73\times 10^{-4}$, which limits the adversary's pulses to $k=495$ for its highest success chance.

\newcommand{\wdththirdsp}{1.725in}
\newcommand{\hghtthirdsp}{\wdththirdsp}

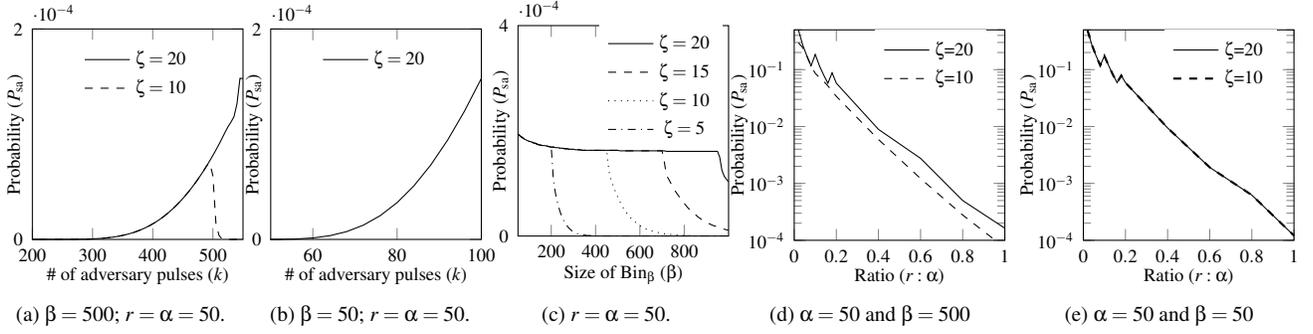
\begin{figure*}
\centering
\subfloat[$\beta=500$; $r=\alpha=50$.]{
\label{khameswa7ed}
\begin{tikzpicture}
\begin{axis}[
width=\wdththirdsp,
height=\hghtthirdsp,
xlabel={\# of adversary pulses ($k$)},
ylabel={Probability ($P_{\text{sa}}$)},
xmin=200, xmax=550,
ymin=0, ymax=0.0002,
ytick ={0,0.0002},
%xtick ={0,100,200},
%extra x ticks={495},
extra x tick style={grid=major},
%extra y ticks={0.35,0.54},
extra y tick style={grid=major,yticklabels={,,}},
%minor x tick num=1,
%grid=minor,
tick label style={font=\scriptsize},
x label style={font=\scriptsize,at={(0.5,0.1)}},
y label style={font=\scriptsize,at={(0.36,0.5)}},
legend columns=1,
legend style={draw=none,at={(0.5,0.95)},anchor=north,}
]
\addplot[] table[col sep=comma]{csv/prob/third/a50b500g20r50.csv};
\addlegendentry{\scriptsize $\zeta=20$}
\addplot[dashed] table[col sep=comma]{csv/prob/third/a50b500g10r50.csv};
\addlegendentry{\scriptsize $\zeta=10$}
%\addplot[dotted] table[col sep=comma]{csv/prob/third/a50b500g5r50.csv};
%\addlegendentry{\scriptsize $\zeta=5$}
\end{axis}
\end{tikzpicture}
}
\hspace{-10pt}
\subfloat[$\beta=50$; $r=\alpha=50$.]{
\label{sadeswa7ed}
\begin{tikzpicture}
\begin{axis}[
width=\wdththirdsp,
height=\hghtthirdsp,
xlabel={\# of adversary pulses ($k$)},
ylabel={Probability ($P_{\text{sa}}$)},
xmin=50, xmax=100,
ymin=0, ymax=0.0002,
ytick ={0,0.0002},
%xtick ={0,100,200},
%extra x ticks={1.5},
extra x tick style={grid=major},
%extra y ticks={0.35,0.54},
extra y tick style={grid=major,yticklabels={,,}},
%minor x tick num=1,
%grid=minor,
tick label style={font=\scriptsize},
x label style={font=\scriptsize,at={(0.5,0.1)}},
y label style={font=\scriptsize,at={(0.36,0.5)}},
legend columns=1,
legend style={draw=none,at={(0.5,0.95)},anchor=north,}
]
\addplot[] table[col sep=comma]{csv/prob/third/a50b50g20r50.csv};
\addlegendentry{\scriptsize $\zeta=20$}
\end{axis}
\end{tikzpicture}
}
%\hspace{\spcbtwn}
\hspace{-15pt}
%\subfloat[Maximum $P_{\text{sa}}$, for all $0\leq k\leq \alpha+\beta$, at $r=\alpha=50$.]{
\subfloat[$r=\alpha=50$.]{
\label{meya}
\begin{tikzpicture}
%\begin{axis}[
%width=3.31in,
%height=\hghtthird,
%ylabel=Rate (ranging/second),
%ytick style={draw=none},
%xmin=0, xmax=200,
%ymin=2000, ymax=4000,
%ytick ={2000,3000,4000},
%%scaled y ticks = true,
%scaled y ticks=base 10:-3,
%every y tick scale label/.style={at={(0.97,1.05)}},
%axis y line*=right,
%axis x line=none,
%y tick label style={font=\scriptsize},
%y label style={font=\scriptsize,at={(1.26,0.5)}},
%legend columns=1,
%legend style={draw=none,at={(0.8,1.15)},anchor=north,}
%]
%\addplot[ultra thick] table[col sep=comma]{csv/prob/third/rate.csv};
%\addlegendentry{\scriptsize Rate}
%\end{axis}
\begin{axis}[
width=\wdththirdsp,
height=\hghtthirdsp,
xlabel=Size of Bin$_\beta$ ($\beta$),
ylabel={Probability ($P_{\text{sa}}$)},
ytick style={draw=none},
xmin=50, xmax=1000,
ymin=0, ymax=0.0004,
xtick ={0,200,400,600,800},
ytick ={0,0.0004},
%extra x ticks={1.5},
extra x tick style={grid=major},
%extra y ticks={0.35,0.54},
extra y tick style={grid=major,yticklabels={,,}},
%minor x tick num=1,
%grid=major,
tick label style={font=\scriptsize},
x label style={font=\scriptsize,at={(0.5,0.1)}},
y label style={font=\scriptsize,at={(0.36,0.5)}},
legend columns=1,
legend style={draw=none,at={(0.68,1.01)},anchor=north,}
]
\addplot[] table[col sep=comma]{csv/prob/third/g20ra.csv};
\addlegendentry{\scriptsize $\zeta=20$}
\addplot[dashed] table[col sep=comma]{csv/prob/third/g15ra.csv};
\addlegendentry{\scriptsize $\zeta=15$}
\addplot[dotted] table[col sep=comma]{csv/prob/third/g10ra.csv};
\addlegendentry{\scriptsize $\zeta=10$}
\addplot[dashdotted] table[col sep=comma]{csv/prob/third/g5ra.csv};
\addlegendentry{\scriptsize $\zeta=5$}
\end{axis}
\end{tikzpicture}
}
%\hspace{\spcbtwn}
\hspace{-10pt}
\subfloat[$\alpha=50$ and $\beta=500$]{
\label{chart:forRbogus}
\begin{tikzpicture}
\begin{axis}[
width=\wdththirdsp,
height=\hghtthirdsp,
xlabel={Ratio ($r:\alpha$)},
ylabel={Probability ($P_{\text{sa}}$)},
xmin=0, xmax=1,
ymin=0.0001, ymax=0.5,
ymode=log,
%ytick ={0,0.25},
%xtick ={0.5,1},
%extra x ticks={1.5},
extra x tick style={grid=major},
%extra y ticks={0.35,0.54},
extra y tick style={grid=major,yticklabels={,,}},
%minor x tick num=1,
%grid=minor,
tick label style={font=\scriptsize},
x label style={font=\scriptsize,at={(0.5,0.1)}},
y label style={font=\scriptsize,at={(0.18,0.5)}},
legend columns=1,
legend style={draw=none,at={(0.65,1)},anchor=north,}
]
\addplot[] table[col sep=comma]{csv/prob/forR/a50b500g20.csv};
\addlegendentry{\scriptsize $\zeta$=20}
\addplot[dashed] table[col sep=comma]{csv/prob/forR/a50b500g10.csv};
\addlegendentry{\scriptsize $\zeta$=10}
\end{axis}
\end{tikzpicture}
}
\hspace{-10pt}
\subfloat[$\alpha=50$ and $\beta=50$]{
\label{chart:forRnobogus}
\begin{tikzpicture}
\begin{axis}[
width=\wdththirdsp,
height=\hghtthirdsp,
xlabel={Ratio ($r:\alpha$)},
ylabel={Probability ($P_{\text{sa}}$)},
xmin=0, xmax=1,
ymin=0.0001, ymax=0.5,
ymode=log,
%ytick ={0,1},
%xtick ={0.5,1},
%extra x ticks={1.5},
extra x tick style={grid=major},
%extra y ticks={0.35,0.54},
extra y tick style={grid=major,yticklabels={,,}},
%minor x tick num=1,
%grid=minor,
tick label style={font=\scriptsize},
x label style={font=\scriptsize,at={(0.5,0.1)}},
y label style={font=\scriptsize,at={(0.18,0.5)}},
legend columns=1,
legend style={draw=none,at={(0.65,1)},anchor=north,}
]
\addplot[] table[col sep=comma]{csv/prob/forR/a50b50g20.csv};
\addlegendentry{\scriptsize $\zeta$=20}
\addplot[thick,dashed] table[col sep=comma]{csv/prob/forR/a50b50g10.csv};
\addlegendentry{\scriptsize $\zeta$=10}
\end{axis}
\end{tikzpicture}
}
\caption{Adversarial success probability in (\ref{eq:finalprobcalc}).}
\label{chart:success}
\end{figure*}

Figure~\ref{meya} shows the effect of $\beta$ on $P_{\text{sa}}$; %Since longer transmission (\ie due to larger $\beta$) will decrease the rate at which two-way ranging can be made, we also plot the rate at the respective value of $\beta$, along the right vertical axis. 
$P_{\text{sa}}$ is almost constant with $\beta$, at around $0.2\times 10^{-3}$, and only starts dropping when $\beta$ is sufficiently large so that the aggregate energy after the adversary's pulses exceeds $\Gamma$. At a certain point, increasing $\beta$ no longer helps. For example, at $\zeta=5$ and $\beta\ge400$, $P_{\text{sa}}\approx0$. $\beta$ should thus be set wisely, reflecting the application's sensitivity to distance increases and channel conditions, to avoid increasing transmission lengths unnecessarily.

\subsubsection{Symbol length ($r$)} 
\label{sec:selectingrsecurity}

\begin{figure}
\centering

\end{figure}

Figures~\ref{chart:forRbogus} and \ref{chart:forRnobogus} plot $P_{\text{sa}}$ against the ratio of $r:\alpha$. As shown, longer symbol length (larger $r$) is better for security; the best results are achieved when the ratio is 1 ($r=\alpha$).

\subsubsection{False positives: noise passing \second{}} 
\label{sec:noisepassing}

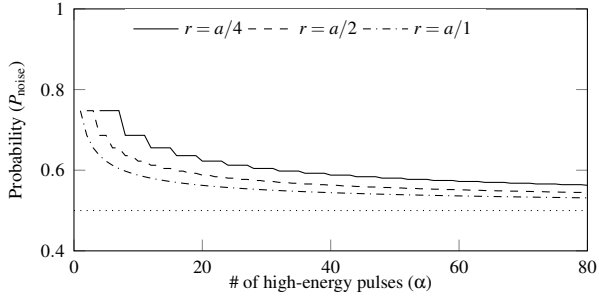
\begin{figure}
\centering
%\subfloat[$\beta=100$]{
%\label{}
\begin{tikzpicture}
\begin{axis}[
width=3.31in,
height=\hghtthird,
xlabel={\# of high-energy pulses ($\alpha$)},
ylabel={Probability ($P_{\text{noise}}$)},
xmin=0, xmax=80,
ymin=0.4, ymax=1,
%ytick ={0,1},
%xtick ={0,100,200},
%extra x ticks={1.5},
extra x tick style={grid=major},
%extra y ticks={0.35,0.54},
extra y tick style={grid=major,yticklabels={,,}},
%minor x tick num=1,
%grid=minor,
tick label style={font=\scriptsize},
x label style={font=\scriptsize,at={(0.5,0.1)}},
y label style={font=\scriptsize,at={(0.07,0.5)}},
legend columns=3,
legend style={draw=none,at={(0.45,1)},anchor=north,}
]
%\addplot[thick] table[col sep=comma]{csv/prob/shortening/b100_kaover2_r1.csv};
%\addlegendentry{\scriptsize $r=1$}
%\addplot[blue] table[col sep=comma]{csv/prob/shortening/b100_kaover2_r2.csv};
%\addlegendentry{\scriptsize $r=2$}
%\addplot[red] table[col sep=comma]{csv/prob/shortening/b100_kaover2_r20.csv};
%\addlegendentry{\scriptsize $r=20$}
\addplot[] table[col sep=comma]{csv/prob/shortening/b100_kaover2_raover4.csv};
\addlegendentry{\scriptsize $r=a/4$}
\addplot[dashed] table[col sep=comma]{csv/prob/shortening/b100_kaover2_raover2.csv};
\addlegendentry{\scriptsize $r=a/2$}
\addplot[dashdotted] table[col sep=comma]{csv/prob/shortening/b100_kaover2_ra.csv};
\addlegendentry{\scriptsize $r=a/1$}
%\addplot[thick] table[col sep=comma]{csv/prob/shortening/alleqala.csv};
%\addlegendentry{\scriptsize $\beta = r = \kappa = \alpha$}

\addplot[dotted,domain=0:120] {0.5};
\end{axis}
\end{tikzpicture}
%}
%\hspace{\spcbtwn}
%\subfloat[$\beta=0$]{
%\label{}
%\begin{tikzpicture}
%\begin{axis}[
%width=\wdth,
%height=\hght,
%xlabel={\# of high-energy pulses ($\alpha$)},
%ylabel={Probability ($P_{\text{noise}}$)},
%xmin=0, xmax=80,
%ymin=0.4, ymax=1,
%%ytick ={0,1},
%%xtick ={0,100,200},
%%extra x ticks={1.5},
%extra x tick style={grid=major},
%%extra y ticks={0.35,0.54},
%extra y tick style={grid=major,yticklabels={,,}},
%%minor x tick num=1,
%%grid=minor,
%tick label style={font=\scriptsize},
%x label style={font=\scriptsize,at={(0.5,0.1)}},
%y label style={font=\scriptsize,at={(0.07,0.5)}},
%legend columns=1,
%legend style={draw=none,at={(0.35,1)},anchor=north,}
%]
%\addplot[dotted] table[col sep=comma]{csv/prob/shortening/b0_kaover2_raover4.csv};
%\addlegendentry{\scriptsize $\kappa = \alpha/2$, $r=a/4$}
%\addplot[dashed] table[col sep=comma]{csv/prob/shortening/b0_kaover2_raover2.csv};
%\addlegendentry{\scriptsize $\kappa = \alpha/2$, $r=a/2$}
%\addplot[] table[col sep=comma]{csv/prob/shortening/b0_kaover2_ra.csv};
%\addlegendentry{\scriptsize $\kappa = \alpha/2$, $r=a$}
%\addplot[dotted,domain=0:120] {0.5};
%\end{axis}
%\end{tikzpicture}
%}
\caption{Probability that noise passes the \second{} check, calculated using (\ref{eq:shorten}); $\kappa = \alpha/2$, $\beta=100$.}
\label{chart:shortening}
\end{figure}

Higher-than-usual noise in the channel might satisfy the \second{} check. Since the receiver backtracks, it is imperative to calculate the probability, $P_{\text{noise}}$, that noise in the channel satisfies that check. Unlike the adversary's pulses targeted to alter the authentic code, such a candidate trail of noise pulses does not get added to the sender's code because they are at different positions. Without loss of generality, we can separate the noise-intervals in low-energy and high-energy, \eg across the median of the distribution of $N^2$. We refer to the number of high-energy intervals as $\kappa$. The probability that noise satisfies the \second{} check is the probability that $x$ of $\kappa$ pulses fell into Bin$_\alpha$, by the probability of satisfying the test in that event, $p'_{\alpha, r}(x)$:
\begin{equation}
\label{eq:shorten}
P_{\text{noise}}(\alpha, \beta, r, \kappa) = \sum_{x=0}^\alpha \left(p'_{\alpha, r}(x) \cdot \frac{\chse{\alpha}{x} \  \chse{\beta}{\kappa-x}}{\chse{\alpha+\beta}{\kappa}} \right)
\end{equation}
where,
\begin{equation}
p'_{\alpha, r}(x) = \sum_{y=0}^r\left(\frac{\chse{\alpha-x}{r-y}\  \chse{x}{y}}{\chse{\alpha}{r}}\cdot \sum_{i=0}^{y} \frac{\chse{\beta-(\kappa-x)}{r-i}\  \chse{\kappa-x}{i}}{\chse{\alpha}{r}}\right)
\end{equation}
This is the probability that an aggregate of $y$ is chosen from Bin$_\alpha$, and of $\leq y$ from Bin$_\beta$. Since we separate along the median, the expected $\kappa$ is $(\alpha+\beta)/2$. Figure~\ref{chart:shortening} plots $P_{\text{noise}}$ against $\alpha$ using (\ref{eq:shorten}) at $\kappa = (\alpha+\beta)/2$ and $\beta=100$. Intuitively (and as the chart confirms), $P_{\text{noise}}\longrightarrow 0.5$ as $\alpha\longrightarrow\infty$.

Since a candidate verification code is discarded as noise if the \second{} check is satisfied with a probability $< P_{\text{noise}}$ (recall: Fig.~\ref{fig:decisiontree}), the adversary must have a success probability of at least $1-P_{\text{noise}}$ to hide the authentic code from the receiver. At $r=\alpha$, $P_{\text{noise}}(80,100,80,40) = 0.53$, and the adversary must thus have a success probability of at least 0.47. As this is much higher than the calculated probabilities in Section~\ref{sec:finalprob}, the adversary will not be able to disguise authentic code as noise. The value 0.53 is a lower-bound; in practice $P_{\text{noise}}$ should be set $\geq 0.53$ depending on applications' requirements and channel conditions.

%!TEX root =  ../main.tex

%\subsection{Simulation Results}
%\label{sec:simulations}

\input{sections/charts/validate}

\subsection{Validating the Probabilistic Model}
\label{sec:validate}
The use of prototype implementation using Software Defined Radios (SDRs) and simulations are well-established methods for evaluating wireless systems. Existing SDRs do not support UWB. Therefore, we validate the probabilistic model above with simulations. The channel condition such as noise, multipath effect, and path loss are important factors to consider while designing a wireless system. The IEEE 802.14.4a~\cite{Molisch_channel_model_CM_5} channel model for different environments is purposefully provided for UWB. The preamble and the verification code are converted into physical layer signals using this model for the outdoor LoS conditions.  The model generates the pulse and multipath components to resemble the real world effect of the channel condition. We assume that upper layers, \eg Medium Access Control (MAC) layer, could decide on when to perform enlargement detection so that it doesn't interfere with other ranging applications. The simulations account for the noise and interference due to the noise figure of the receiver and multipath components.	To verify the simulation setup, we performed a thorough evaluation to cross-check simulation metrics with previous proof-of-concept implementation \cite{mridula_eprint_UWB_PR}. Each pulse uses 500 MHz bandwidth, and the sampling time between consecutive pulses is 1 $\mu s$. Transmission power is limited to -35 dBm/MHz, well under the limits applied by the FCC/ETSI regulations~\cite{fontana2007observations}. The energy is further reduced to adapt to path loss model and extra losses ($E$; \cf~Fig.~\ref{fig:pathloss}).

An adversary is simulated to inject $k$ signals to annihilate or distort the authentic code, and to replay a delayed and amplified versions of the authentic signals. Similar to our assumptions, the adversary in the simulator is capable of annihilating the pulse and its multipath if the phase is guessed correctly; it doubles the amplitude of the pulse otherwise. The time difference between authentic and delayed signals is $\delta = 200 ns$ in the simulations (see Fig.~\ref{fig:main}).

Before demodulation, additive white Gaussian noise (AWGN) is added to the signal. The receiver in Section~\ref{sec:symbol_receiver} is implemented for code verification; it always locks on to the highest peak, \ie the peak generated by the adversary due to its replay attack. The communication range is considered $100m$, and the backtracking restricted to $660ns$.

The goal of our validation is to (1) confirm the probabilistic model's correctness, and (2) analyze the effect of the parameters abstracted from the model, namely noise and the receiver's ability to reconstruct the signal after long distance propagation. In practice, the latter point can be accounted for by increasing the number of pulses ($n=\alpha+\beta$)---see below.

{\bf Validating $P_{b\beta>b\alpha}$.} 
Figure~\ref{chart:validatebox} shows the validation for $P_{b\beta>b\alpha}$, at a simulated distance between both devices of $d=10 m$. A boxplot is drawn at distinct $k$, where each scenario is run $10^6$ times. The results confirm that abstracting noise from the model does not largely affect its accuracy. Next we show the effect of longer distances on the model.%, and how that can be accounted for.

{\bf Validating $P_{\text{sa}}$.} Figure~\ref{fig:All_together} shows the validation for $P_{\text{sa}}$, at $r = \alpha$ and $P_{noise} = 0.8$. Results are shown for different $k$, at distances of $10m$ and $100m$. Each scenario is run $10^6$ times, and $P_{\text{sa}}$ is calculated as the proportion of these where the adversary succeeded to hide the authentic code. 
Again the results show comparable patterns between the model and simulations. There is a slight horizontal shift at $k$ due to the abstracted noise. In the simulator, $\Gamma$ is set as in~(\ref{eq:Gamma}), which may be a bit too high or low depending on actual noise patterns. In Fig.~\ref{fig:All_together_first}, $\Gamma$ was relatively low, causing a drop in the simulated $P_{\text{sa}}$ at smaller $k$ compared to the model. In Fig.~\ref{fig:All_together_second}, $\Gamma$ was relatively high, replicating $P_{\text{sa}}$ at higher $k$. 

Another difference between simulations and the model manifests with increasing the distance $d$ between both devices. In practice, in UWB, receivers increase their ability to reconstruct the signals (hence, the SNR) by aggregating over more pulses. We noticed that the model provides such comparable probability patters when we decrease $\alpha$ and $\beta$ in the model proportionally with increasing $d$ in simulations. For example in Fig.~\ref{fig:All_together_second} where $d=100m$, $\alpha$ and $\beta$ in the simulator had to be increased from 15 and 158 to 50 and 500 respectively ($\sim$ tripled) to account for the increased distance. 

{\bf Validating the false positives.} We also used simulations to confirm that noise would not be falsely mistaken for authentic code upon proper selection of $P_\text{noise}$ and $\Gamma$. For various distances between $10m$ and $100m$, the probability of a false positive was $\sim 1\times 10^{-6}$, confirming the noise analysis in Section~\ref{sec:noisepassing}.

$~$

In conclusion, the simulated probabilities follow comparable patterns with the model, and are in the same range. The model derived herein thus serves as a formal means for evaluating the efficacy and suitability of UWB-ED in practice. The results also show that the channel condition, such as path loss, noise, and interference due to multipath components, does not affect the performance and security of the system. An adversary can increase the noise level, which can increase false positives. High false positives may eventually cause DoS (which the adversary can mount anyway by jamming the channel), but the adversary remains unable to enlarge distances.

%Assumed:
%------
%for d=0.01, alpha = alpha/1.11850585 and beta = beta/1.11850585		??
%for d=0.1, alpha = alpha/1.452605 and beta = beta/1.452605		??
%for d=1, alpha = alpha/1.8865 and beta = beta/1.8865		??

%Fact:
%------
%for d=10, alpha = alpha/2.45 and beta = beta/2.45
%for d=100, alpha = alpha/3.15 and beta = beta/3.15

\begin{figure}
\centering
\subfloat[]{
\label{fig:All_together_first}
\begin{tikzpicture}
\begin{axis}[
width=\wdththirdsp,
height=\hghtthirdsp,
xlabel={\# of adversary pulses ($k$)},
ylabel={Probability ($P_{\text{sa}}$)},
xmin=0, xmax=1,
ymin=0, ymax=0.00015,
%ytick ={0,0.25},
%xtick ={0,100,200},
%extra x ticks={1.5},
extra x tick style={grid=major},
%extra y ticks={0.35,0.54},
extra y tick style={grid=major,yticklabels={,,}},
%minor x tick num=1,
%grid=minor,
tick label style={font=\scriptsize},
x label style={font=\scriptsize,at={(0.5,0.1)}},
y label style={font=\scriptsize,at={(0.22,0.5)}},
legend columns=1,
legend style={draw=none,at={(0.5,1.44)},anchor=north,}
]
\addplot[line width=0.15mm] table[col sep=comma]{csv/prob/validate/a20b204g5r20.csv};
\addlegendentry{\scriptsize Prob. Model: $\alpha = 20$, $\beta = 204$}
\addplot[line width=0.1mm,dashed] coordinates {
(0/550,0)
(50/550,0)
(100/550,0)
(150/550,0.0001)
(200/550,0)
(250/550,0)
(300/550,0)
(350/550,0)
(400/550,0)
(450/550,0)
(500/550,0)
};
\addlegendentry{\scriptsize Sim: $d=10$, $\alpha = 50$, $\beta = 500$}
%\zeta = 3;  0, 0, 0, 0, 0, 0, 0, 0, 0, 0, 0
%\zeta = 5; 0, 0, 0,.0001, 0, 0, 0, 0, 0, 0, 0
\end{axis}
\end{tikzpicture}
}
\hspace{-13pt}
\subfloat[]{
\label{fig:All_together_second}
\begin{tikzpicture}
\begin{axis}[
width=\wdththirdsp,
height=\hghtthirdsp,
xlabel={\# of adversary pulses ($k$)},
ylabel={Probability ($P_{\text{sa}}$)},
xmin=0, xmax=1,
ymin=0, ymax=6e-4,
%ytick ={0,0.25},
%xtick ={0,100,200},
%extra x ticks={1.5},
extra x tick style={grid=major},
%extra y ticks={0.35,0.54},
extra y tick style={grid=major,yticklabels={,,}},
%minor x tick num=1,
%grid=minor,
tick label style={font=\scriptsize},
x label style={font=\scriptsize,at={(0.5,0.1)}},
y label style={font=\scriptsize,at={(0.25,0.5)}},
legend columns=1,
legend style={draw=none,at={(0.5,1.44)},anchor=north,}
]
\addplot[line width=0.15mm] table[col sep=comma]{csv/prob/validate/a15b158g5r15.csv};
\addlegendentry{\scriptsize Prob. Model: $\alpha = 15$, $\beta = 158$}
\addplot[line width=0.1mm,dashed] coordinates {
(50/550,0)
(100/550,0)
(150/550,0)
(200/550,0)
(250/550,0)
(300/550,0.0002)
(350/550,0.0001)
(400/550,0.0001)
(450/550,0.0005)
(500/550,0.0002)
(550/550,0.0004)
};
\addlegendentry{\scriptsize Sim: $d=100$, $\alpha = 50$, $\beta = 500$}
%\zeta = 3; 0, 0, 0, 0, 0, 0, 0, 0, 0, 0, 0
%\zeta = 5; 0,0,0,0,0,.0002,0.0001,0.0001,0.0005,0.0002,0.0004
\end{axis}
\end{tikzpicture}
}
\caption{The attack is detected when the aggregate energy is between $\gamma$ and $\Gamma$, but $P_{b\beta>b\alpha}$ is more than $P_{\text{noise}}$. The attack is also detected when energy aggregate is more than $\Gamma$; $\zeta=5$.}
\label{fig:All_together}
\end{figure}

\section{Discussion}
\label{sec:security}

{\bf Adaptive attacks.} An adversary can notice the effect of each of its added pulses on the resultant energy, whether annihilated or amplified. It can then adapt its attack strategy by dynamically deciding $k$ based on the number of pulses it has added/annihilated so far during the transmission. The adversary can then utilize its knowledge of $n$, $\alpha$ and $\beta$ in order to, not only decide the optimal value of $k$ statically before the transmission begins, but also adjust their distribution in realtime. This attack does not succeed because the adversary cannot control the resultant pulse phase. Injecting excessive energy in Bin$_\beta$ exceeds $\Gamma$; injecting in Bin$_\alpha$ does not guarantee annihilation because of the unpredictable phase.

{\bf Varying energy levels.} To achieve perfect signal annihilation, an adversary uses the same amplitude expected at the receiver. Instead of injecting $k$ pulses each with a constant energy of, \eg $2\mu W$, the adversary can inject one pulse with an energy of, \eg $2k\mu W$. If all $k$ pulses fell in Bin$_\beta$, the aggregate energy would be the same as when that single high-energy pulse also falls in Bin$_\beta$. However, intuitively, the adversary is better off injecting multiple pulses with constant energies for two reasons. First, multiple pulses in Bin$_\beta$ have higher chances of being selected than a single pulse, thus evading the \second{} check. Second, for those that fall in Bin$_\alpha$, any leftover energy after annihilating a pulse, regardless of the phase, will be counted towards the overall aggregate, thus hurts the adversary's cause.

{\bf Influencing $\Gamma$ through distance shortening.} Instead of enlarging distances directly, the adversary can first mount a distance-reduction attack to trick the devices into using higher $\Gamma$ (recall: smaller signal attenuation due to shorter path loss leads to higher $\Gamma$ calibration). It is thus imperative to complement UWB-ED with a distance-reduction detection~\cite{Brands1994,DB_EPFL,mridula_eprint_UWB_PR}. Devices should alternate between both techniques; \eg if distances of $d_1$ and $d_2$ are verified using respectively UWB-ED and a distance-reduction detection technique, it should be concluded that the actual distance, $d$, is in the range $d_1 \leq d \leq d_2$ ($d_1$ is a lower bound, $d_2$ an upper). 

{\bf Influencing the number of pulses, $n$.} An adversary can inject a low stream of noise-like energy, not too high to be detected as jamming. However because $\Gamma$ is set beforehand, it is not influenced by the adversary. %The energy in the channel, irrespective of the source, is compared to $\Gamma$. 
By injecting noise, the adversary actually hurts its own cause as it reduces the amount of energy it can use strategically to prevent code detection. 

{\bf Integrating UWB-ED with 802.15.4z and 5G.} The 802.15.4z enhanced impulse radio task group is defining a series of physical layer improvements to provide secure and precise ranging \cite{802.15.4z}. Those include additional coding, preambles, and improvement to existing modulations to increase ranging integrity and accuracy. UWB-ED is a potential candidate for enlargement detection in 802.15.4z. It adheres to the low pulse repetition (LRF) mode frequency (1-2 MHz), works with non-coherent receivers, and supports up to $100m$.

The 3GPP technical specifications groups are designing the 5G-new radio technology, and it aims to include secure and precise ranging based on wireless signals~\cite{5G-NR-Ericsson,5G_Vehicular_Networks_Positioning}. Properties such as high carrier frequencies, large bandwidths, large antenna arrays, device-to-device communication, and ultra-dense networking will help attain this objective. It is early to say the exact modulation techniques 5G will use for distance measurement, but it is safe to assume that wideband will be used to attain position accuracy; beamforming techniques will achieve long distances. This system is equivalent to setting $r=1$ herein without restrictions on $\alpha$, as transmission power restrictions imposed on UWB do not apply to 5G. However, the security of 5G can be increased further, as it allows for the use of beamforming and coherent receivers.

\section{Related Work}
\label{sec:relatedwork}

Detecting enlargement attacks has lately been a prominent research area. Previous literature explored timing acquisition at the preamble, and data ambiguity at payload. Taponecco~\etal~\cite{Taponecco_Overshadow} show that the success of enlargement attacks using replay (or overshadowing) depends on the amount of delay the adversary introduces. Such success is harder for controllable attacks, where the adversary is required to position nodes at specific locations. Compagno~\etal~\cite{Multiple_leading_edge} provide a probabilistic model for the success of overshadowing attacks, which captures different channel conditions and leading edge detection techniques for ToA estimation. None of the above efforts considered adversarial signal annihilation.

Tippenhauer~\etal~\cite{Nils_Integrity} explored a theoretical approach to detect adversarial signal annihilation for distance enlargement: using a single pulse-per-symbol (consecutive integration windows represent a symbol). They found that modulation with a 2ns slot size, \ie mostly equivalent to a pulse width, might help detect signal annihilation. This, however limits the ranging technique to short distances. The effect of multipath on that scheme in practice is also unclear, since reflected signals would directly interfere with authentic ones causing distortion (no empty gaps between authentic pulses). In contrast, UWB-ED allows for increased distances by increasing the symbol length, and the sampling time between consecutive pulses is sufficient to handle the multipath effect.

\section{Conclusion}
\label{sec:conc}

We present UWB-ED---the first known technique to detect distance-enlargement attacks against standard UWB ranging systems. UWB-ED is readily deployable for current off-the-shelf receivers, requiring no additional infrastructure. Evaluation is performed by deriving the probability of adversarial success in mounting distance enlargement attacks. Results show that the verification code structure herein prevents signal annihilation. The code also allows the use of longer symbol length at the receiver, which is essential to achieve longer distance in the energy constrained UWB system. UWB-ED is thus a good candidate for enlargement detection in practice (\eg for 802.15.4z and 5G).

% you can also use the wonderful epsfig package...
%\begin{figure}[t]
%\begin{center}
%\begin{picture}(300,150)(0,200)
%\put(-15,-30){\special{psfile = fig1.ps hscale = 50 vscale = 50}}
%\end{picture}\\
%\end{center}
%\caption{Wonderful Flowchart}
%\end{figure}

\bibliographystyle{plain}
\bibliography{bib/main}

%{\footnotesize \bibliographystyle{acm}
%\bibliography{IEEEabrv,bib/bibliography}}

%
%\IfFileExists{\jobname.ent}{
%   \theendnotes
%}{
%   %no endnotes
%}
%\appendix
%\input{sections/appendix}

\begin{figure*}
	\centering
	\includegraphics[width=1\linewidth]{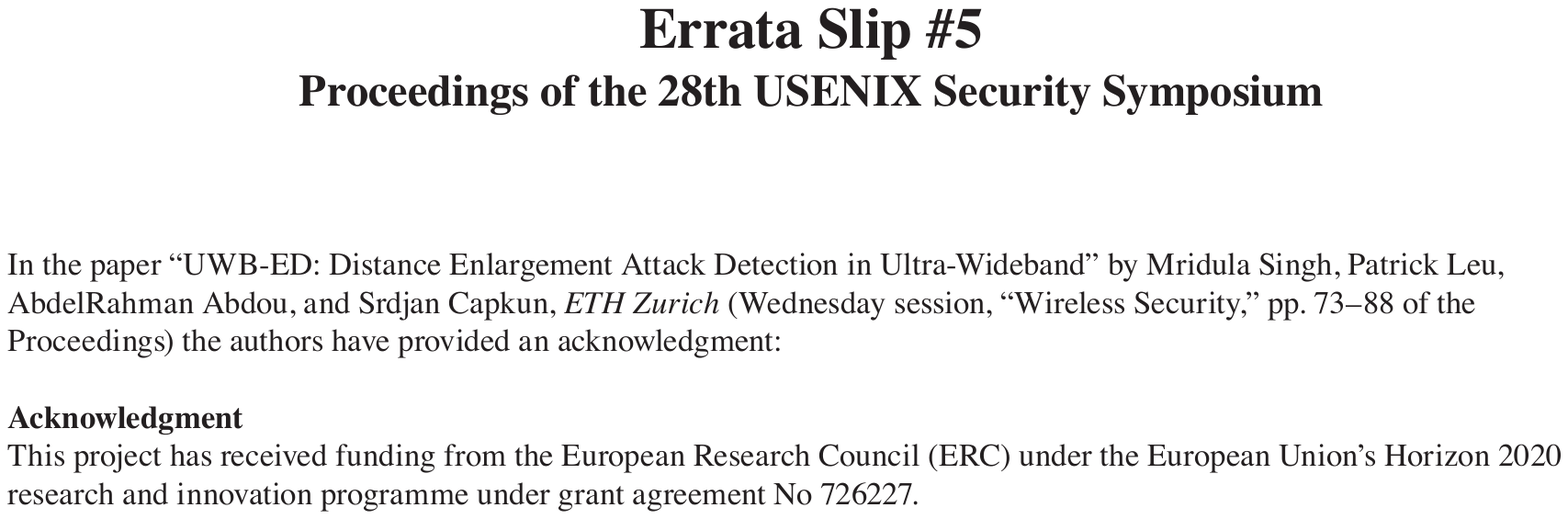}
	\label{fig:uwbeddd}
\end{figure*}

\end{document}